\documentclass[preprint,10pt]{elsarticle}

\usepackage{amsmath, amsthm, amssymb,amssymb,graphicx,fancyhdr,fancyhdr,helvet,ifpdf,setspace,graphics}
\usepackage{booktabs}
\pdfoutput=1
\usepackage{pstricks}
\usepackage{cancel}
\usepackage{fullpage}
\usepackage{enumerate}
\usepackage[normalsize]{subfigure}
\usepackage[section]{placeins}
\journal{Journal Name}

\begin{document}
\begin{frontmatter}
\title{Large Eddy Simulation of Combined Wind-wave Loading on Offshore Wind Turbines}

 \author[lsu]{Tianqi Ma}
 \author[lsu]{Chao Sun\corref{cor}}
 \ead{csun@lsu.edu}
 \cortext[cor]{Corresponding author}
 \address[lsu]{Department of Civil and Environmental Engineering, Louisiana State University, Baton Rouge, Louisiana 70803, USA}


\begin{abstract}

Wind-wave interactions impose wind forcing on wave surface and wave effects on turbulent wind structures, which essentially influences the wind-wave loading on structures. Existing research treats the wind and wave loading separately and ignores their interactions. The present study aims to characterize the turbulent airflow over wave surfaces and wave dynamics under wind forcing and analyze the coupled wind-ave loading on offshore wind turbines. A high-fidelity two-phase model is developed to simulate highly turbulent wind-wave fields based on the open-source program OpenFOAM. A numerical case study is conducted to simulate extreme wind-wave conditions, where coupled wind-wave fields are applied. Simulation result shows that the weighted region of turbulence depends on the relative speed between wind velocity and wave phase speed. The intensity of wave induced turbulences and the height of wave influenced region are affected by the wind velocity and wave heights. Higher wind velocities induce greater turbulence, which can by increased by over 100$\%$. Then the combined wind-wave loading on offshore wind turbines is simulated under operational and extreme conditions. Under operational conditions, the wind-wave coupling effect on the combined loading is minimal. However, under extreme conditions, the coupled wind-wave fields lead to an increase in the average aerodynamic loading and a significant amplification of the fluctuation in the aerodynamic loading. Specifically, the maximum bending moment $F_y$ at both the tower bottom and the monopile bottom experiences an increase of around $6\%$. Furthermore, the wind-wave coupling effect is evident in the standard deviation of the aerodynamic loading at the tower bottom. The standard deviation of the shear force at the tower bottom increases by up to $45\%$. Also, the standard deviation of the bending moment at the tower bottom increases by approximately $27\%$. This study reveals the importance of considering the wind-wave coupling effect under extreme conditions, which provides valuable insights into the planning and design of offshore wind turbines.

\end{abstract}


\end{frontmatter}

\section{Introduction}

\label{S:1}

The increasing demand for clean and renewable energy sources has spurred interest in harnessing the immense potential of offshore wind energy. Offshore wind turbines, also known as wind farms, have emerged as a promising solution to meet the growing global energy needs. These turbines capitalize on the offshore environment, characterized by high wind speeds and unobstructed airflows, enabling the installation of larger turbines and enhanced energy production. However, the deployment of offshore wind turbines presents significant engineering challenges due to the requirement of withstanding harsh marine conditions, including high wind speeds and extreme weather events. In particular, the occurrence of strong tropical cyclones brings about extreme winds, surges, and waves that can cause extensive damage to offshore structures.

Traditionally, engineering practices have treated wind and wave loading on offshore wind turbines as separate phenomena. Wind fields are typically modeled using spectrum or atmosphere boundary layer models, while wave fields are described using wave potential theories \cite{sun2018bi,li2018wind,doubrawa2019load}. Some studies have focused on wind loading on monopile-supported offshore wind turbines without considering the effects of currents and waves \cite{feyzollahzadeh2016wind}. Others have investigated the long-term bending moment on monopile-supported wind turbines, considering the influence of wind speed and wave height \cite{rendon2014long}. Oh \textit{et al.} proposed a method to generate hydrodynamic loads for finite element analysis of a wind energy conversion system with a monopile foundation for an offshore wind turbine without considering wind \cite{oh2013preliminary}. However, these studies do not capture the interaction between wind and waves. In reality, the interaction between wind and waves alters the characteristics of the wind-wave flow field, affecting the combined wind-wave pressures on structures. Wave breaking, in particular, generates vorticity and turbulence, intensifying the exchange of momentum and energy between wind and waves. Previous studies have examined the impact of wave breaking on structures, but have mainly focused on the isolated effects of wave breaking, neglecting the combined wind and wave loadings \cite{winter2020tsunami,wienke2005breaking}.

During high-category hurricanes, wind and waves become coupled through the exchange of momentum and heat flux at the air-water interface. As a result, the wind-wave interaction modifies the characteristics of the wind-wave flow field and influences the combined wind-wave pressures on structures. Accurately assessing the environmental load effects on offshore wind turbines is crucial, particularly when they are subjected to concurrent wind, wave, surges, and currents. Several researchers have emphasized the importance of investigating the environmental load effects on offshore wind turbines when subjected to current wind, wave, and current conditions \cite{bobillier2001physical,buljac2022concurrent}. However, many experimental studies conducted on this topic have been limited to small-scale laboratory experiments, which may underestimate the influence of aerodynamic forces.

Apart from experimental methods, numerical techniques have been employed to analyze the combined wind and wave loading on offshore structures. Traditional CFD analysis simplifies the problem by assuming a uniform wind, neglecting the strong inherent turbulence in the wind above the waves. To accurately replicate realistic wind turbulence conditions, spectral methods are commonly used to generate turbulent wind fields. Some studies have utilized CFD and turbulence models, such as the Mann wind turbulence model, to analyze bottom-fixed wind turbines \cite{li2015coupled}. These studies have shown that turbulence increases wake diffusion. Previous research has also explored the impact of turbulent wind and wind shear on floating offshore wind turbine (FOWT) structures through the application of CFD methods. However, these studies have primarily examined specific wind inflow conditions and have not fully analyzed the changes in wind structure over waves, especially under extreme wind and wave conditions \cite{zhou2022exploring}.

In summary, existing studies on offshore wind turbines typically consider wind and wave loading separately and do not account for their combined effects. Experimental investigations into the combined wind-wave-current loading on structures have mostly been conducted using small-scale laboratory testing, which may underestimate the influence of aerodynamic forces. Numerical methods have primarily focused on uniform wind loading and have neglected the inherent strong turbulence in the wind above the air-water interface, resulting in inadequate quantification of the combined wind-wave pressures on structures subjected to extreme wind-wave conditions. For the wind loading, the turbulent aerodynamic loading is not fully understood. For the wave loading, most existing literature is focused on tsunami-like waves and extreme breaking waves. As a result, hurricane induced wave forces which are essentially contributed by shallow water nonlinear wave effects are not well characterized. In addition, the wind and waves are fully coupled, turbulent high winds induced wind-driven effects that significantly influence the evolution and breaking of waves, which yields the maximum wave pressures, are not covered. The wave profiles also change the wind structures and induce turbulence near the interface, which also have influence on the wind peak pressure and the aerodynamic pressure on the structures. Therefore, it is crucial to comprehend the interactions between wind and waves, particularly under extreme conditions, to gain a comprehensive understanding of the combined wind and wave loading on structures.

To address this knowledge gap, the present study aims to characterize the turbulent air fields above air-water surfaces and their coupling effects on a real-scale monopile-supported offshore wind turbine. A high-fidelity two-phase flow model is developed to simulate the highly turbulent coupled wind-wave flow fields. This model is then applied to analyze the flow field characteristics of extreme wind and wave conditions on a scale of $10^2$ m. The coupled wind-wave fields are further utilized to investigate the combined wind and wave loading on offshore structures under operational and shut-off conditions, considering both successful and failed pitch control. The operational condition represents the normal operation of the wind turbine actively generating power, while the shut-down condition with successful pitch control corresponds to intentional shutdown with adjusted blade positions to minimize wind loadings. The shut-off condition with failed pitch control represents a scenario where the blade pitch control mechanism malfunctions, leading to inefficient control of wind loadings. The remainder of the paper is organized as follows: Section 2 presents a two-phase model for highly turbulent wind-wave flow fields, Section 3 describes the post-processing method for analyzing complex wind structures above waves, Section 4 characterizes the coupled wind and wave fields under extreme conditions, Section 5 analyzes the coupled wind-wave loading and separate wind-wave loading on wind turbines, and finally, Section 6 provides the conclusions of the study.

\section{Model Description}
\label{S:2}

\subsection{Governing Equations}
The water and air are assumed as incompressible flow. The mass conservation and Navier-Stokes equations are

\begin{equation}
\label{eq:11}
\nabla \cdot \mathbf{U} = 0
\end{equation}

\begin{equation}
\label{eq:22}
 \frac{\partial{\rho \mathbf{U}}}{\partial{t}} + \nabla \cdot (\rho \mathbf{UU}) = -\nabla p_{d} - \mathbf{g}\cdot\mathbf{x}\nabla{\rho} +\nabla \cdot (\mu_{eff}\nabla \mathbf{U})+ \nabla{\mathbf{U}}\cdot\nabla{\mu_{eff}} + \sigma\kappa\nabla{\alpha}
\end{equation}
where $\mathbf{U}$ is the velocity vector; $\rho$ is the fluid density; $p_{d}=p-\rho\mathbf{g}\cdot \mathbf{x}$ is the dynamic pressure by subtracting the hydrostatic part from total pressure $p$; $\mathbf{g}$ is the gravitational acceleration; $\mathbf{x}$ is the position vector; $\mu_{eff}=\rho(\nu+\nu_t)$ is the effective dynamic viscosity, in which $\nu$ and $\nu_t$ are the kinematic and turbulent eddy viscosity respectively. $\nu_t$ is modeled by a turbulence model. $\sigma\kappa\nabla{\alpha}$ is a surface tension term where $\sigma$ is the surface tension coefficient, $\kappa$ is the free surface curvature, and $\alpha$ is the volume fraction. In this study, $\sigma$ is set as 0.07 N/m. The volume of fluid (VOF) method is used to mark the air-water interface. An indicator scalar function $\alpha = \frac{V_{water}}{V_{cell}}$ is used to represent the fractional volume of a cell occupied by water. For a two-phase air-water flow, $\alpha=1$ represents that the cell is full of water, $\alpha=0$ represents the cell is full of air, and $0<\alpha<1$ indicates that the cell contains the free surface. Then the density $\rho$ and viscosity $\mu$ of fluid at an arbitrary location can be written as
\begin{equation}
\label{eq:33}
\rho = \alpha\rho_{water}+(1-\alpha)\rho_{air}
\end{equation}

\begin{equation}
\label{eq:44}
\mu = \alpha\mu_{water}+(1-\alpha)\mu_{air}
\end{equation}
where $\rho_{water}$ and $\rho_{air}$ represent water density and air density respectively; $\mu_{water}$ and $\mu_{air}$ are water viscosity and air viscosity. With adopting the VOF method, the two immiscible fluids (water and air) are modeled as an effective continuous flow and are governed by one set of conservation equations. The advection of the indicator function is governed by the transport equation:

\begin{equation}
\label{eq:55}
\frac{\partial\alpha}{\partial t}+\nabla\cdot(\mathbf{U}\alpha)+\nabla\cdot[\mathbf{U_r}\alpha(1-\alpha)]=0
\end{equation}
where a compression term $\nabla\cdot[\mathbf{U_r}\alpha(1-\alpha)]$ is introduced to reduce the interface smearing effect. The compression term is acting only in the free surface zone because of the inclusion of $(1-\alpha)\alpha$. $\mathbf{U_r}$ is a compression velocity.

The turbulent kinetic viscosity $\nu_t$ is modeled by a turbulence model. In this study, the large eddy simulation (LES) method is adopted with large scale eddies resolved and small scale turbulences modeled with the subgrid-scale stress (SGS) model. A wall adapting local eddy viscosity (WALE) is adopted to model the subgrid-scale stress, which can model accurate wall boundary layer\cite{nicoud1999subgrid,ben2017assessment,weickert2010investigation}. The WALE model calculates the eddy viscosity based on the invariants of the velocity gradients.

\begin{equation}
\label{eq:66}
\nu_t = (C_w\Delta)^2\frac{(S_{ij}^{d}S_{ij}^d)^{3/2}}{(\bar S_{ij}\bar S_{ij})^{5/2}+(S_{ij}^d S_{ij}^d)^{5/4}}
\end{equation}
where $\bar S_{ij}=1/2(\partial \bar u_i/\partial x_j+\partial \bar u_j/\partial x_i)$ is the rate-of-strain tensor; $\bar{( )}$ represents filtered variable; parameter $S_{ij}^d$ is the traceless symmetric part of the square of the velocity gradient tensor, which is calculated using Eqn. (\ref{eq:77}).

\begin{equation}
\label{eq:77}
S_{ij}^d = \frac{1}{2}\left(\frac{\partial \bar u_k}{\partial x_i}\frac{\partial \bar u_j}{\partial x_k}+\frac{\partial \bar u_k}{\partial x_j}\frac{\partial \bar u_i}{\partial x_k}\right)-\frac{1}{3}\delta_{ij}\frac{\partial \bar u_k}{\partial x_l}\frac{\partial \bar u_l}{\partial x_k}
\end{equation}
where $\delta _{ij}$ is the Kronecker’s delta.

\subsection{Boundary conditions}
\subsubsection{Wave wind boundary conditions at inlet boundary}
In the present study, the wave is generated at the inlet boundary through specifying the free surface elevation and water particle velocities, which is implemented referring to the olaFlow framework developed by Higuera et al.\cite{olaFlow}. Different waves, linear wave and non-linear waves, can be generated at the boundary using the classic wave theories. Similarly, the wind velocity is prescribed for the air phase at the inlet boundary. Uniform or other (log or power) mean wind profiles can be generated above the water surface, where the wind velocity in the air-water interface grid is consistent with the water velocity at wave surface. In addition to a mean velocity, the wind field at the inlet is prescribed as the summation of the mean flow and turbulent fluctuations. To generate the turbulent wind, a turbulent spot method is implemented to generate spatially and temporally correlated turbulent fluctuations, which possesses prescribed Reynolds stress and integral length and satisfies continuity constraint \cite{kroger2018generation,poletto2013new}. In turbulent spots method, a set of turbulent spots are randomly distributed at the inlet boundary and are convected by the mean velocity through the boundary. For the $i^{th}$ spot, an inner velocity distribution is set as $u_n^i(\mathbf{r}-\mathbf{r}^i)=\varepsilon_{n}^i f_n(\mathbf{r}-\mathbf{r}^i)$, where n is the component number, $\mathbf{r}^i$ is the center of the spot, $\varepsilon_{n}^i$ is a uniformly distributed random number between -1 and 1. The inner velocity distribution determines spectra and the integral length scales. The velocity fluctuations at point $\mathbf{r}$ are the sum of contributions from all spots $\mathbf{u}=\sum_{i=1}^N{\mathbf{u}^i(\mathbf{r}-\mathbf{r}^i)}$. This method can generate anisotropic turbulence through introducing the anistropy into the turbulent spots \cite{kornev2007synthesis}. The velocities at the inlet boundary is fixed as specified values with wave particle velocities applied for the water phase and wind velocities for the air phase. The pressure condition at the inlet boundary is set as zero gradient.

\subsubsection{Wave absorption}
A wave relaxation zone is added near the outlet boundary to avoid the reflection of waves from outlet boundaries. Based on \cite{jacobsen2012wave}, an explicit relaxation method, $\phi=(1-w_R)\phi_{target}+w_R\phi_{computed}$, is applied to correct the indicator scalar function $\alpha$ and velocity $\mathbf{U}$, where $\phi$ represents $\alpha$ or $\mathbf{U}$, $w_R$ is a weighting function of local coordinate system in the relaxation region. Referring to \cite{fuhrman2006numerical}, an exponential weighted distribution is selected. In this study, the relaxation technique is only effective for waves and has no effect on airflow through setting the weight function as zero in the air phase. The outlet boundary is treated as pressure boundary conditions with the total pressure condition set as fixed values and the velocity condition set as zero gradient.  

\subsection{Monopile Wind Turbine Model}

An NREL 5MW offshore monopile wind turbine is selected. The monopile is fully modeled, while the tower and wind blade use the advanced actuator line turbine model implemented in NREL SOWFA \cite{churchfield2017advanced,troldborg2009actuator}. The actuator line method offers a more efficient and computationally economical approach compared to fully resolving the blades. It involves projecting the computed lift and drag forces along a one-dimensional actuator line onto the three-dimensional computational fluid dynamics (CFD) mesh as a body force, which is imposed to the Navier-Stokes equations.

\begin{equation}
\label{eq:revise1}
\vec{f}_{turb}\left(x,y,z \right) =\vec{F}g\left(x,y,z\right) 
\end{equation}
where $\vec{f}_{turb}$ is the projected body force, $\vec{F}$ is the actuator line element force, $g(r)$ is the force projection function, which integrates to one. The force at an actuator point comprises an orthogonal lift component to the local velocity and a parallel parasite drag component.

\begin{equation}
\label{eq:revise2}
\vec{F}=-\rho \frac{c}{2}\hat{U}\left( C_l\vec{e}_z\times \vec{U}+C_{dp}\vec{U} \right) 
\end{equation}
where $\vec{e}_z$ represents the vector aligned with the wing or blade axis, $C_l$ and $C_{dp}$ are the lift and parasite drag coefficients, respectively. $\vec{U}$ is the local velocity of the fluid relative to the blade, $\hat{U}$ is the magnitude of the freestream velocity $\vec{U}_{\infty}$, $c$ corresponds to the chord length of the blade. The advanced actuator line model projects the line forces using the following projection function.

\begin{equation}
\label{eq:revise3}
g_{AAL}(x_c,x_t,x_r)=\frac{1}{\varepsilon_c\varepsilon_t\varepsilon_r\pi^{3/2}}\exp\left[-\frac{(x_c-x_{c,0})^2}{\varepsilon_c^2}--\frac{(x_t-x_{t,0})^2}{\varepsilon_t^2}-\frac{(x_r-x_{r,0})^2}{\varepsilon_r^2}\right]
\end{equation}
where $x_c$, $x_t$, and $x_r$ represent the coordinates in the chord-wise, thickness-wise, and radial directions, respectively. The subscript '0' indicates the reference location for applying the Gaussian. The Gaussian widths in each direction are denoted as $\varepsilon_c$, $\varepsilon_t$, and $\varepsilon_r$. The projected body force is then added to the momentum equation. 

\subsection{Numerical Method}

In the present study, the simulations are performed using open-source library OpenFOAM (Open-source Field Operations And Manipulations)\cite{weller1998tensorial}. A standard solver interFoam is provided in OpenFOAM for incompressible two fluid flows, which is based on the Finite Volume Method and VOF surface capturing method \cite{habchi2013partitioned,holzmann2016mathematics}.

\section{Post-process}
\label{S:3}

To analyze the wind-wave coupling interaction, the wave elevation and wind-wave velocity field are sampled during the CFD simulation process. 

\subsection{Wave elevation}
To obtain the surface elevation, several wave gauges are defined along the numerical wave tank. At each gauge location, a linear distribution of points are defined in the vertical direction to obtain the value of $\alpha$ through linear interpolation of the simulated results at the cell center. The water elevation is calculated using the sampled $\alpha$ values:

\begin{equation}
\label{eq:88}
\eta = \sum_{i=1}^{N-1}{(h_{i+1}-h_i)\frac{\alpha_{i+1}+\alpha_{i}}{2}}+h_{min}
\end{equation}
where $h$ is the height of the sampled points; $h_{min}$ is the minimum height of the points along the vertical line.

\subsection{Coordinate transformation and phase detection}
\label{S:3_2}
In the numerical simulation, wind wave velocities are obtained in Cartesian coordinates, with $x$, $y$, and $z$ representing the streamwise, spanwise, and vertical coordinates, respectively. The velocity components $u$, $v$ and $w$ are functions of ($x$, $y$, $z$). To analyze the air flow above the wave surface, a wave surface following coordinate is introduced. The wind-wave fields are transformed to a orthogonal co-ordinate system as proposed in Ref. \cite{benjamin1959shearing}. The wave is a combination of a series of Fourier wave components. The wave elevation ($\eta$) can be expressed as:
\begin{equation}
\label{eq:1}
\eta \left( x,t \right) =\sum_n{a_ne^{i\left(k_nx-2\omega_nt+\phi _n \right)}}
\end{equation}
where $a_{n}$, $\omega_{n}$, $k_{n}$, $\phi_{n}$ are the amplitude, circular frequency, wave number, and phase of the $n^{th}$ mode. A coordinate system which follows the wave surface is introduced, where the orthogonal co-ordinate($\xi$, $\zeta$) is represented by Cartesian coordinates ($x$, $z$) as:
\begin{eqnarray}
\label{eq:2}
\xi \left( x,z \right) &=& x-i\sum_n{a_ne^{i\left( k_n\xi -2\omega_nt+\phi _n \right)}e^{-k_n\zeta}} \nonumber\\
\zeta \left( x,z \right) &=& z-\sum_n{a_ne^{i\left( k_n\xi -2\omega_nt+\phi _n \right)}e^{-k_n\zeta}}
\end{eqnarray}

In the orthogonal coordinate, the wave surface corresponds to $\zeta =0$. For air turbulence statistics above wave surface, a phase average approach is used to quantify the statistical properties of turbulence in order to study the interaction between wind and surface waves. With the calculated wave elevation $\eta$, the wave phase of the wind-wave fields can be obtained by applying a Hilbert transform to the wave profiles. A wave-phase decomposition is then applied to the wind-wave velocity field to analyze the air structure above the wave surface. A quantity $q$ can be decomposed into the sum of a phase-averaged quantity $<q>$ and a turbulent quantity $q'$.
\begin{equation}
\label{eq:4}
q(x,y,z,t)=<q>(\xi,\zeta)+q'(x,y,z,t)
\end{equation}

With wave phase calculated using Hilbert transform to wave profiles, the phase averaged quantity ($<q>(\xi,\zeta)$) can be obtained and the turbulent quantity $q'$ can be obtained by subtracting $<q>$ from $q$. The turbulent quantity $q'$ is used to represent the turbulence of wind or wave velocities. Also, the phase-averaged quantity $<q>$ can be further decomposed into a phase-independent total averaged quantity $\bar{q}$ across all phases and a wave-coherent quantity $\tilde{q}$. Via this decomposing method, the instantaneous 3D turbulence flow field $q(x,y,z,t)$ can be expressed as:

\begin{equation}
\label{eq:5}
q(x,y,z,t)=\bar{q}(\zeta)+\tilde{q}(\xi,\zeta)+q'(x,y,z,t)
\end{equation}

\section{Numerical case study}

\begin{figure}[h!]

\centering\includegraphics[width=0.5\linewidth]{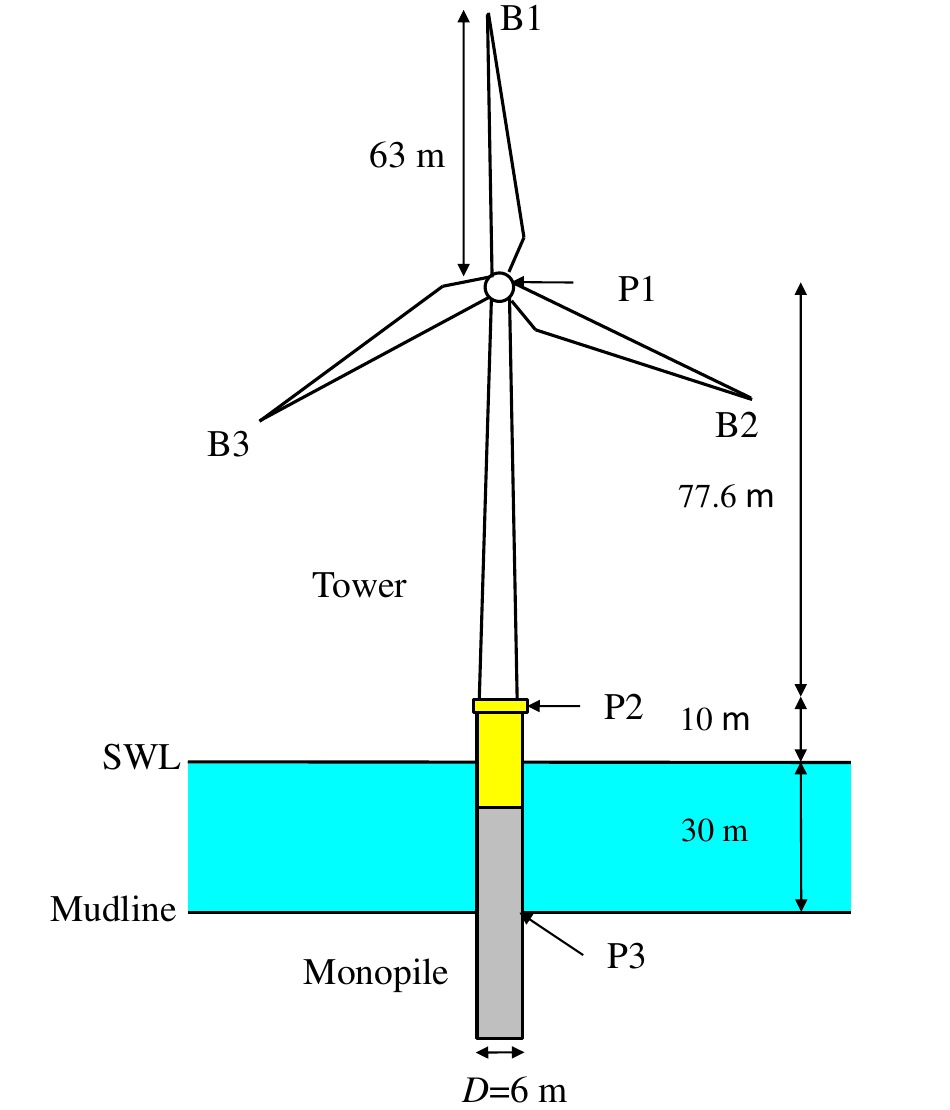}.
\caption{Schematic of NREL 5MW offshore wind turbine}
\label{fig:MonopileOWT}
\end{figure}

In this section, the numerical model is applied to analyze the characteristics of coupling wind-wave fields and the combined wind and wave loading on a monopile wind turbine. The NREL 5MW wind turbine is analyzed with a rotor diameter of 126 m and a hub height of 87.6 m above the still water line (SWL). The schematic and structure dimensions of the offshore wind turbines and monopile foundation are shown in Fig. \ref{fig:MonopileOWT}. The computational domain has a size of 1000 m $\times$ 400 m $\times$ 300 m with a water depth of 30 m. Initially, the mesh resolution is 4 m $\times$ 4 m $\times$ 1 m. In order to capture details near the still water surface within the region $15 m<z<75m$, the mesh is refined twice. The refined mesh in this region has a resolution of 1 m $\times$ 1 m $\times$ 0.25 m. To accurately resolve the wind field in proximity to the wind turbine blades and towers, the meshes are further refined in these specific areas. The monopile and the transient piece, shown as the yellow and grey parts in Fig. \ref{fig:MonopileOWT}, are meshed in the simulation domain. The total number of meshes of the computational domain is around 48 million. The flow velocity at the monopile surface and the transient piece is set as zero. The tower and the blades are not meshed and modeled using actuator lines models to project the line forces to the 3D body forces, which are applied as source terms to the flow fields. Two wind wave conditions are simulated: operational and extreme cut-off states. When a turbine is cut off during extreme winds, the blade pitch is typically adjusted to align with the wind direction, aiming to minimize the aerodynamic loadings.  However, in real applications, the pitch controller may not function properly, failing to maintain the pitch at the desired 90-degree angle. In this study, a pitch angles of $60^{\circ}$ and $0^{\circ}$ are also simulated to analyze the behavior of the wind turbine under extreme conditions. The wind wave conditions and wind turbine conditions are summarised in Table \ref{table:WindTurbineCondition}. For the four wind-wave cases mentioned in Table \ref{table:WindTurbineCondition}, simulations are performed for a monopile wind turbine under two scenarios: coupled wind-wave fields and separated wind and wave fields. Specifically, for each wind-wave condition, simulations are conducted for the monopile wind turbine under three configurations: (1) coupled wind-wave fields, (2) sole wind fields with wave height set to zero, and (3) sole wave fields with wind speed set to zero. The wind loading in the sole wind fields and the wave loading in the sole wave fields are individually calculated. The combination of these two loadings represents the wind-wave loading acting on structures under separate wind and wave fields. By comparing the separate wind-wave loading with the combined wind-wave loading under coupled wind-wave fields, we can analyze the influence of the coupled wind-wave effect on the structures.

\begin{table}[h]
\centering
\caption{Parameters for wind and wave conditions and wind turbine conditions}
\label{table:WindTurbineCondition}
\begin{tabular}{l l| l l l |l l l l | l l}
\hline
\multicolumn{2}{c|}{Case} & \multicolumn{3}{c|}{Logarithmic wind conditions} & \multicolumn{4}{c|}{Wave conditions} & \multicolumn{2}{c}{Wind turbine conditions}\\
\hline
{No.} &\textbf{$C_{p}/u_{*}$} & \textbf{$U_{10}$}&\textbf{$z_0$}&\textbf{$u_{*}$}& type&\textbf{$a$}&\textbf{$\lambda$}&\textbf{$T$}&{rotor speed}&{pitch}\\
& &(m/s) & (cm) &(m/s)& &(m)&(m)&(s)& & \\
\hline
1 & 32.21  & 10 & 0.02 & 0.379 & stokes I& 2.00 & 100 & 8.19 & controlled & controlled\\
2 & 6.25 & 55 & 0.02 & 2.084 & stokes III & 5.00 & 106.6 & 8.19& 0 & 90\\
3 & 6.25 & 55 & 0.02 & 2.084 & stokes III & 5.00 & 106.6  & 8.19& 0 & 60\\
4 & 6.25 & 55 & 0.02 & 2.084 & stokes III & 5.00 & 106.6  & 8.19& 0 & 0\\
\hline
\end{tabular}
\end{table}

As mentioned before, the present study simulates two wind-wave scenarios: operational and extreme conditions. At the inlet boundary, the water phase velocity is defined through the prescribed regular wave and the air phase velocity is defined as turbulent wind speed, using the corresponding wind and wave parameters listed in Table \ref{table:WindTurbineCondition}. Based on the wave parameters, stokes linear and nonlinear waves are generated in operational and extreme conditions, separately. The parameters for generating the wind turbulences, such as the integral length scales, turbulence intensities, and Reynolds stress, are determined referring to the ESDU 85020 \cite{esdu2001characteristics}. A fixed shear stress ($\tau = \rho u_*^2$) is applied to maintain the inlet profiles over a long distance. Simulations for all the cases are carried out for 200 s in parallel with the entire domain divided into 512 subdomains and the maximum CFL number set as 0.5. Statistics are obtained after 80 s with a time interval of 0.01 s.

\subsection{Wind structure over waves}
\label{s:windEvolutionAlongWave}

\begin{figure}[h!]

\centering\includegraphics[width=0.9\linewidth]{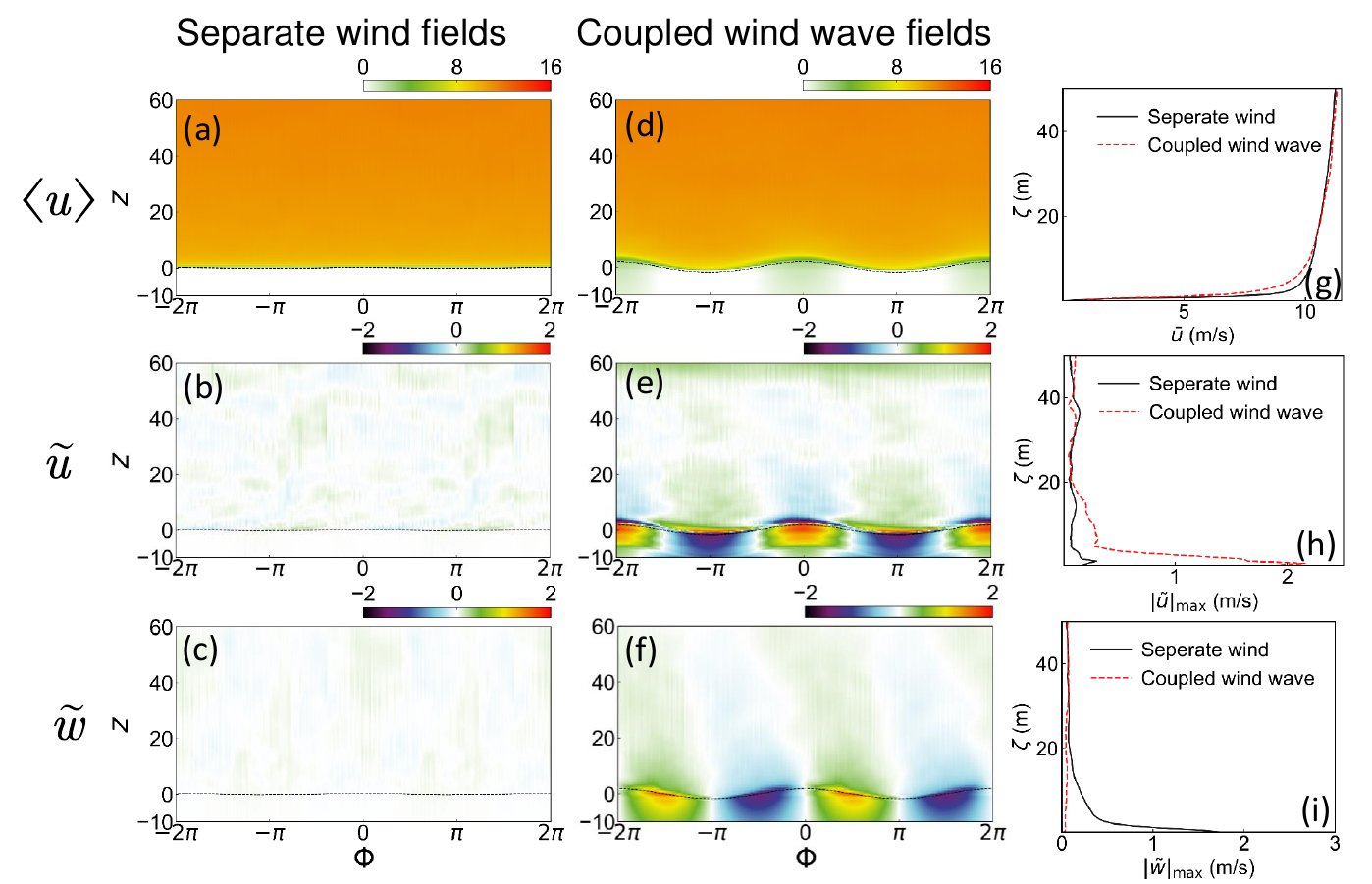}.
\caption{Phase averaged velocity field in case 1 with $U_{10}$ = 10 m/s and a=2 m. (a,d) Phase averaged velocity $<u>$; (b,e) Wave-coherent streamwise velocity $\tilde{u}$; (c,f) Wave-coherent vertical velocity $\tilde{w}$ (g) Vertical profile of average streamwise velocity across all phases $\bar{u}$; (h) Vertical profile of maximum wave-coherent streamwise velocity across all phases $|\tilde{u}|_{max}$. (i) Vertical profile of maximum wave-coherent vertical velocity across all phases $|\tilde{w}|_{max}$.}

\label{fig:OperationalCase_WaveInduce}
\end{figure}

First, the characteristics of the coupled wind and wave fields are analyzed and compared with those of seperate wind fields. To study the interaction between turbulent winds and surface waves, the phase-average method presented in Section \ref{S:3_2} is used to quantify the statistical properties of wind turbulence. The phase averaged streamwise velocity $<u>$, wave coherent velocities $\tilde{u}$ and $\tilde{w}$ of coupled and separate wind wave fields at $x=400$ m in operational conditions (case 1) are presented in the first and second columns in Fig. \ref{fig:OperationalCase_WaveInduce}. The wave coherent velocities ($\tilde{u}$) represent the variance of the phase averaged wind velocity with wave phase ($\tilde{u}=<u>-\overline{u}$). Over still water surface, the phase averaged velocity $<u>$ has no change with wave phase, as shown in Fig. \ref{fig:OperationalCase_WaveInduce}(a), and the wave coherent velocities $\tilde{u}$ and $\tilde{w}$ approach zero, as shown in Fig. \ref{fig:OperationalCase_WaveInduce}(b) and Fig. \ref{fig:OperationalCase_WaveInduce}(c). Over wave surface with a wave height of 4 m, the wave coherent velocity $\tilde{u}$ is positive above troughs and negative above crests near the wave surface. The wave coherent velocity $\tilde{w}$ close to the wave surface matches well with the water particle velocities, with negative $\tilde{w}$ at the upwind side of waves and positive at the downwind side. The intensified regions of wave coherent velocities are very close to the wave surfaces. In the case of wind travelling over old waves with a wave age of 32.2, the wave coherent velocities $\tilde{u}$ and $\tilde{w}$ are mainly induced by wave moving. Fig. \ref{fig:OperationalCase_WaveInduce}(g) shows the vertical averaged velocity $\bar{u}$ in the separate wind fields and coupled wind wave fields. The vertical averaged quantities are calculated by averaging following the vertical coordinate $\zeta$. The black lines denote the results in the separate wind fields and the red dashed lines denote the results in the coupled wind wave fields. The vertical averaged velocity $\bar{u}$ in the coupled wind wave fields is close to that in the separate wind fields except that very close to the wave surface below the height of 10 m. The maximum absolute wave coherent velocities across all phases ($|\tilde{u}|_{max}$, $|\tilde{w}|_{max}$) are shown in Fig. \ref{fig:OperationalCase_WaveInduce}(h) and Fig. \ref{fig:OperationalCase_WaveInduce}(i). In the operational case with $a=2$ m and $U_{10}=10$ m/s, the wave induces wave coherent velocities to 2.1 m/s in the streamwise direction and 1.7 m/s in the vertical direction. The maximum wave coherent velocities $|\tilde{u}|_{max}$ and $|\tilde{w}|_{max}$ are smaller than 1 m/s ($0.1U_{10}$) above heights of 2.7 m and 1.5 m, respectively, which indicates that the wave profiles influence the averaged wind fields $\tilde{u}$ and $\tilde{w}$ to heights of 2.7 m and 1.5 m.

\begin{figure}[h!]

\centering\includegraphics[width=0.9\linewidth]{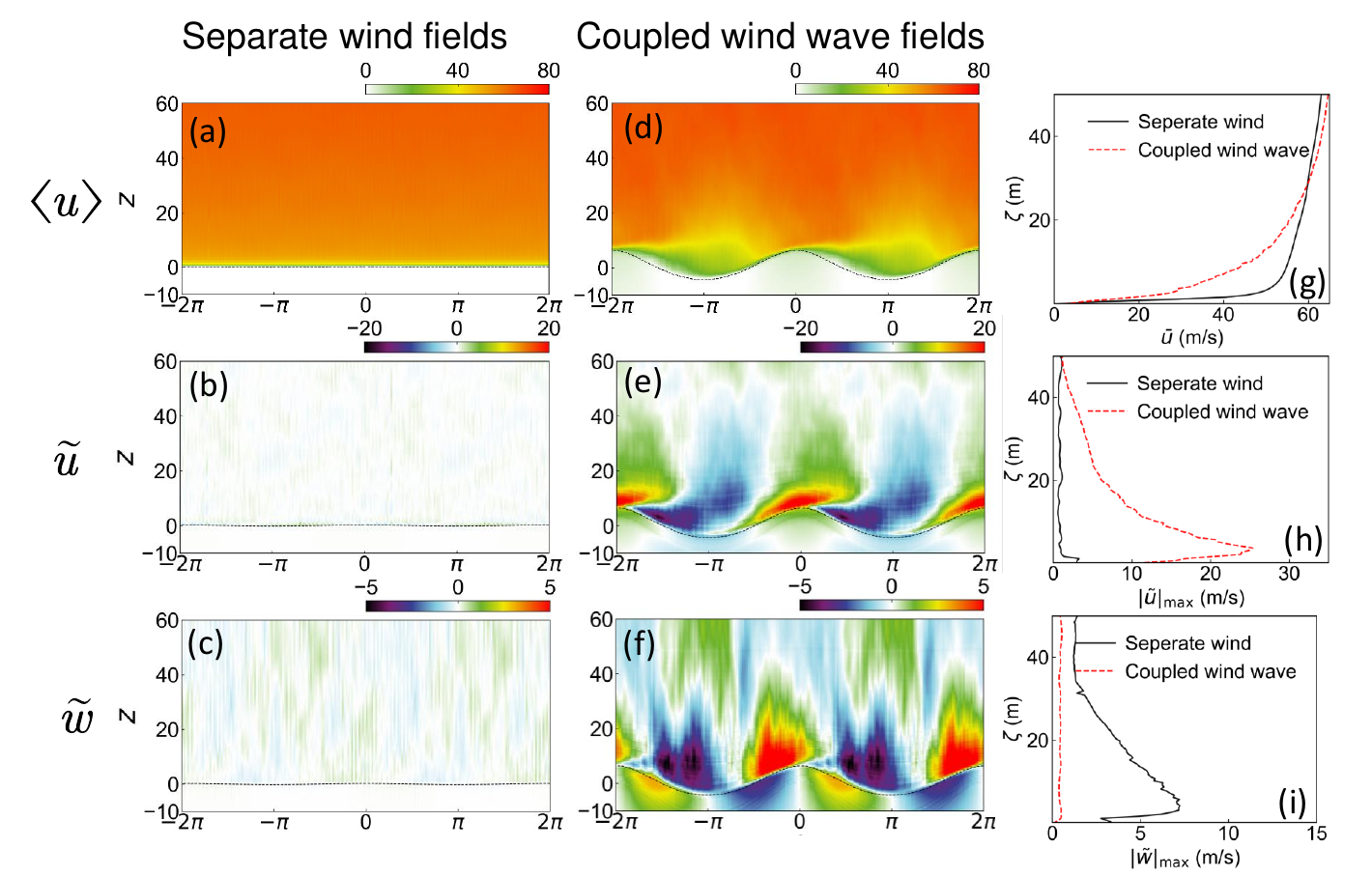}.
\caption{Phase averaged velocity field in case 2 with $U_{10}$ = 55 m/s and a=5 m. (a,d) Phase averaged velocity $<u>$; (b,e) Wave-coherent streamwise velocity $\tilde{u}$; (c,f) Wave-coherent vertical velocity $\tilde{w}$ (g) Vertical profile of average streamwise velocity across all phases $\bar{u}$; (h) Vertical profile of maximum wave-coherent streamwise velocity across all phases $|\tilde{u}|_{max}$. (i) Vertical profile of maximum wave-coherent vertical velocity across all phases $|\tilde{w}|_{max}$.}

\label{fig:ExtremeCase_WaveInduce}
\end{figure}

Fig. \ref{fig:ExtremeCase_WaveInduce} shows the averaged wind fields at $x=400$ m in extreme conditions. The results in the separate wind fields are shown in the first column and the results in the coupled wind wave fields are shown in the second column. The vertical profiles are shown in the third column. Comparing the phase averaged velocity $<u>$ in the separate wind fields (Fig.\ref{fig:ExtremeCase_WaveInduce}(a)) and that in coupled wind wave fields(Fig.\ref{fig:ExtremeCase_WaveInduce}(d)), one can find that the wave decreases the averaged streamwise velocity downwind of the waves, which is caused by the sheltering effect. Near the wave surface, the wave coherent velocity $\tilde{u}$ of coupled wind wave fields is positive along upwind surface and negative along the downwind surface. As height increases, the $\tilde{u}$ contours 
tilts downwind. The wave coherent velocity $\tilde{w}$ is positive at the upwind face and negative at the downwind face. Comparing to the $\tilde{w}$ contour under wave surface, the $\tilde{w}$ contour above wave surface exhibits an opposite negative-positive pattern. The magnitude of $\tilde{u}$ in wind fields is larger than that in the water, which indicates that the wave coherent velocity $\tilde{u}$ in extreme condition is caused by the fast moving wind above the waves. The wave coherent velocities are dominant near the wave surface and negligible away from the surface. Under extreme condition with $a=5$ m and $U_{10}=55$ m/s, the difference of the vertical averaged velocity $<u>$ between separate wind and coupled wind wave fields is evident under the height of 20 m, as shown in Fig. \ref{fig:ExtremeCase_WaveInduce}(g). The wave induces wave coherent velocities to 22.8 m/s in the streamwise direction and 5.6 m/s in the vertical direction. The wave coherent velocity $\tilde{u}$ and $\tilde{w}$ are intensified ($0.1U_{10}$) close to the wave surfaces up to a height of 18 m and 5.5 m separately. 

\begin{figure}[h!]

\centering\includegraphics[width=0.9\linewidth]{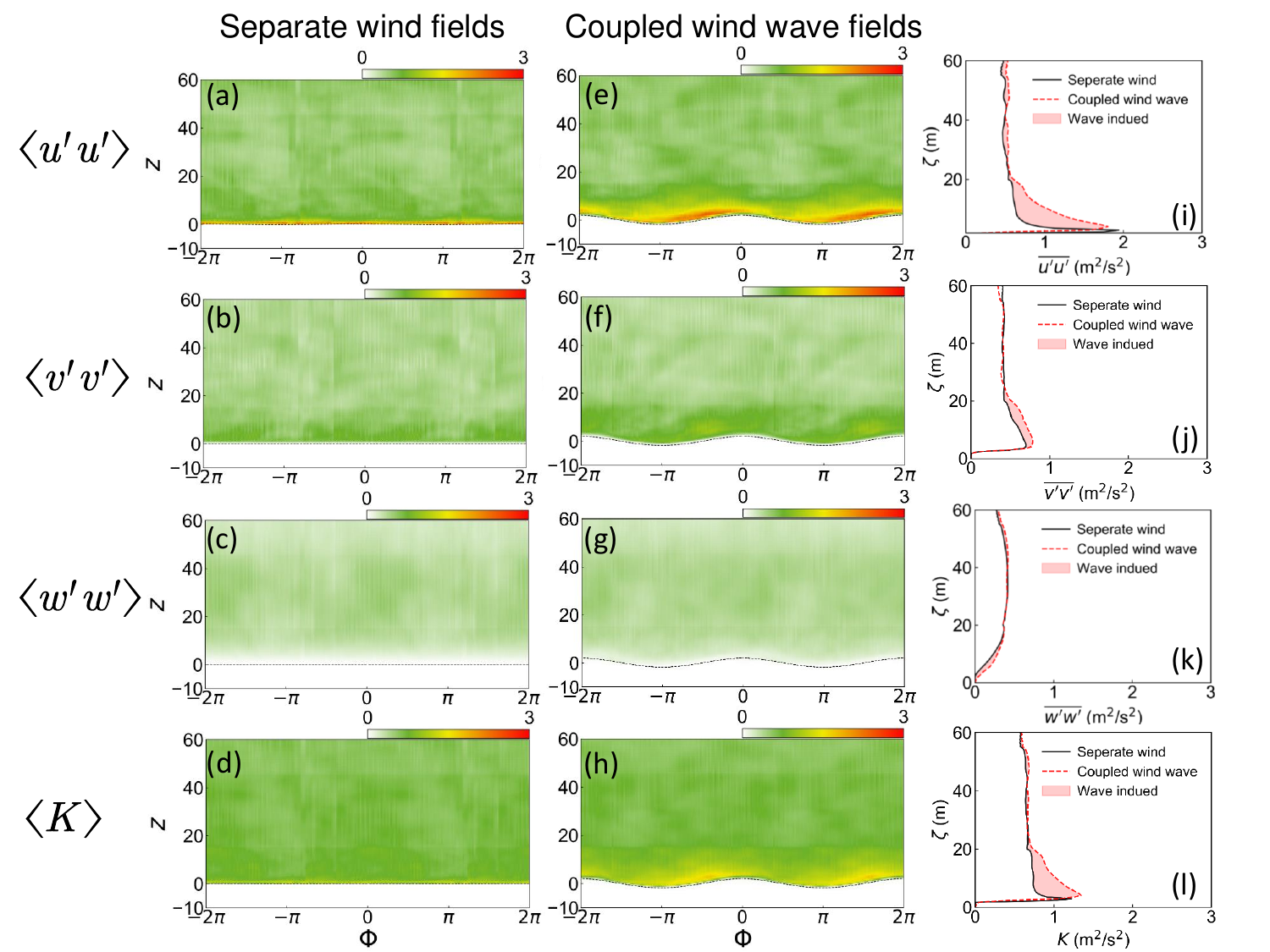}.
\caption{Phase averaged turbulence in case 1 with $U_{10}$ = 10 m/s and a=2 m. (a,e) Phase averaged turbulent variance $<u'u'>$; (b,f) Phase averaged turbulent variance $<v'v'>$; (c,g) Phase average turbulent variance $<w'w'>$; (d,h) Phase averaged turbulence kinetic energy $<K>$; Total mean (across all phases) turbulent variances (i) $\overline{u'u'}$, (j) $\overline{v'v'}$, and (k) $\overline{w'w'}$; (l) Total mean turbulence kinetic energy $\overline{K}$.}

\label{fig:OperationalCase_turbulence}
\end{figure}

In addition to its effect on the averaged wind fields, the wave surface profiles influence the wind turbulence. Turbulent variances ($<u'u'>$, $<v'v'>$, and $<w'w'>$), and turbulence kinetic energy ($<K>$) of separate wind fields and coupled wind wave fields at $x=400$ m in operational condition are presented in the first and second column in Fig. \ref{fig:OperationalCase_turbulence}. Comparison between the Fig. \ref{fig:OperationalCase_turbulence} (a) and Fig. \ref{fig:OperationalCase_turbulence} (e) shows that the wave enhances wind turbulence near the wave surface. In case 1 with $U_{10}$ = 10 m/s and $a$ = 2 m, the regions with intense $<u'u'>$ are located at the upwind side of the waves, as shown in the Fig. \ref{fig:OperationalCase_turbulence}(e). In case 1, with the wave age $C_{p}/u_{*}$ = 32.21, the wind blows over fast moving old waves. Near the wave surfaces, the waves travel faster than the winds and the sheltering effect takes place at the upwind faces of the waves. Through comparing the Figs \ref{fig:OperationalCase_turbulence}(f) and (g) and Figs \ref{fig:OperationalCase_turbulence}(b) and (c), one can find that the variances ($<v'v'>$, and $<w'w'>$) in coupled wind wave fields are close to that in separated wind fields. The phase averaged turbulence kinetic energy $<K>$ has a similar pattern as that of $<u'u'>$ because $<K>$ is mainly contributed by the variances $<u'u'>$ in the stream-wise direction. To quantify the wave induced turbulence, the vertical averaged variances ($\overline{u'u'}$, $\overline{v'v'}$, and $\overline{w'w'}$) and vertical averaged turbulence kinetic energy ($\overline{K}$) for all phases are shown in the third column. Comparing the turbulence in separated wind fields and in coupled wind wave fields indicates that the waves induce turbulence near the wave surface which is marked in red. As the wind velocity $U_{10}$ is close to the wave phase velocity, the wave-generated wind turbulence is tiny in case 1, as shown in Fig. \ref{fig:OperationalCase_turbulence}(l). At a height of 10 m, the vertical averaged turbulence kinetic energy $\overline{K}$ slightly increases from 0.75 m$^2$/s$^2$ in separate wind fields to 1.0 m$^2$/s$^2$ in coupled wind wave fields. The intensified region of turbulence kinetic energy is within the height of 20 m.

\begin{figure}[h!]

\centering\includegraphics[width=0.9\linewidth]{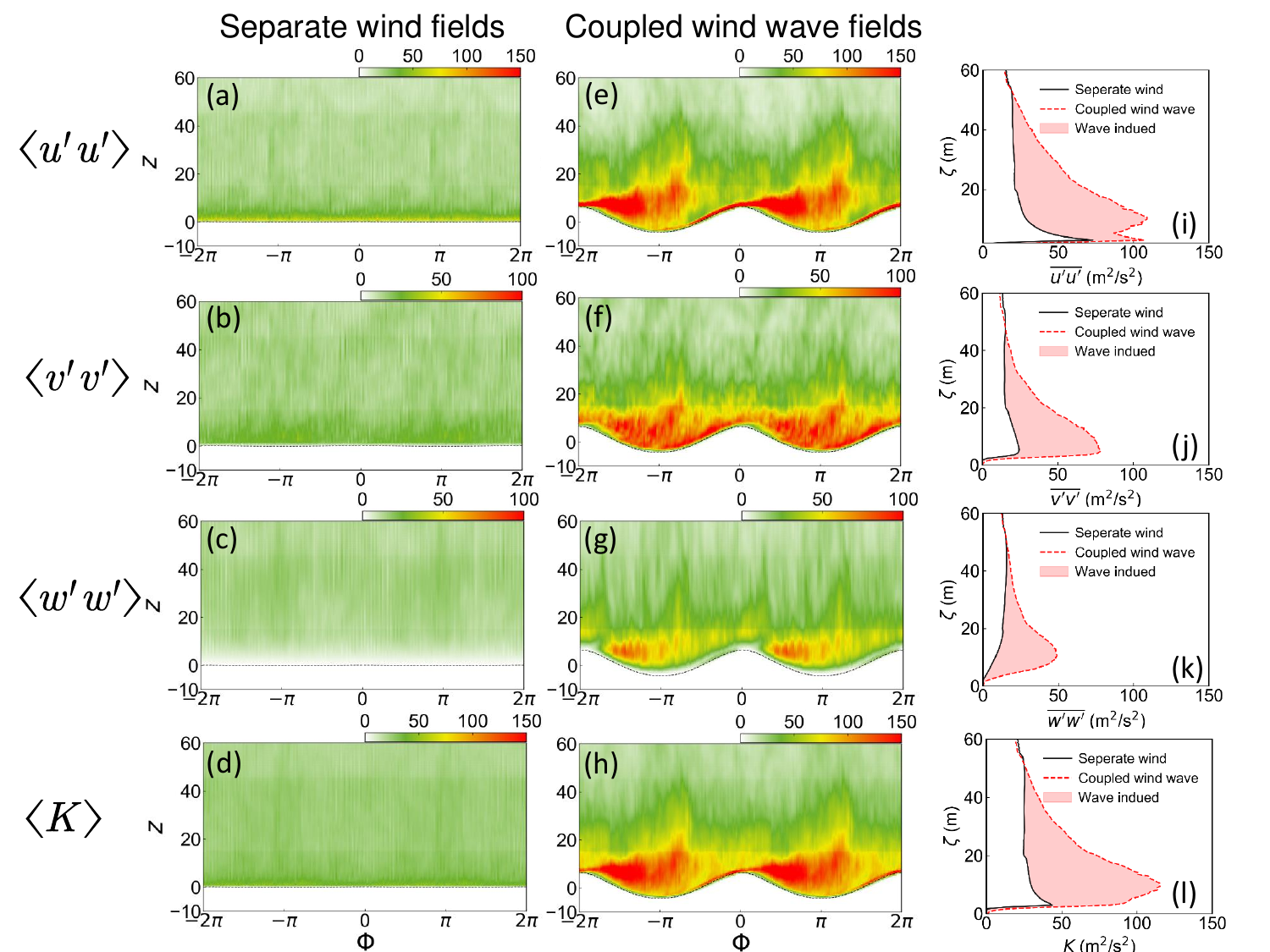}.
\caption{Phase averaged turbulence in case 2 with $U_{10}$ = 55 m/s and a=5 m. (a,e) Phase averaged turbulent variance $<u'u'>$; (b,f) Phase averaged turbulent variance $<v'v'>$; (c,g) Phase average turbulent variance $<w'w'>$; (d,h) Phase averaged turbulence kinetic energy $<K>$; Total mean (across all phases) turbulent variances (i) $\overline{u'u'}$, (j) $\overline{v'v'}$, and (k) $\overline{w'w'}$; (l) Total mean turbulence kinetic energy $\overline{K}$.}

\label{fig:ExtremeCase_turbulence}
\end{figure}

The wind turbulence in extreme conditions is shown in Fig. \ref{fig:ExtremeCase_turbulence}. Comparing the phase averaged turbulent variances ($<u'u'>$, $<v'v'>$, and $<w'w'>$) of coupled wind wave fields and that of separate wind fields, one can find that the wave significantly enhances wind turbulence near the wave surface in extreme conditions. An intense turbulent variance $<u'u'>$ region can be observed past the crest of the waves. The regions with maximum variances, $<u'u'>$, happen at the downwind faces of waves. The $<v'v'>$ and $<w'w'>$ are also intensified in the extreme condition. The phase averaged turbulence kinetic energy $<K>$ has a similar pattern as that of $<u'u'>$, which is also observed in the operation condition because $<K>$ is mainly contributed by the variances $<u'u'>$. In the operational case with $U_{10}$ = 10 m/s and $a$ = 2 m, the regions with intense turbulent kinetic energy are located at the upwind side of the waves, as shown in the Fig. \ref{fig:OperationalCase_turbulence}(h), which is opposite to that in the extreme condition with the intense turbulent $<K_t>$ regions locate at the downwind side of the waves, as shown in Fig. \ref{fig:ExtremeCase_turbulence}(h). The different pattern in Fig. \ref{fig:OperationalCase_turbulence}(h) is caused by a "reversed sheltering effect" \cite{buckley2016structure}. In the extreme condition, Figs. \ref{fig:ExtremeCase_turbulence}(i), (j), (k), and (l) show that, at a height of 10 m, the streamwise variance ($\overline{u'u'}$) increases from 28 m$^2$/s$^2$ to 108 m$^2$/s$^2$, the spanwise component increase from 22 m$^2$/s$^2$ to 73 m$^2$/s$^2$, the vertical component increase from 8 m$^2$/s$^2$ to 49 m$^2$/s$^2$ and the total averaged turbulence kinetic energy $\overline{K}$ increases from 29 m$^2$/s$^2$ to 115 m$^2$/s$^2$. The wave strengthened the wind turbulence by more than 100$\%$. The waves influence the wind fields dramatically near the wave surface in extreme conditions and increase the wind turbulence up to a height of 50 m above the wave surfaces. Through comparison between case 1 and case 2, one can find that higher wave heights and stronger wind forcing induce stronger turbulences and affect the wind field to a higher region. 

\subsection{Application in offshore wind turbines}

The previous subsection indicates that wave surfaces influence the wind fields by inducing turbulences and changing the structure of averaged wind fields. In this section, the simulated coupled wind wave fields are utilized to analyze the combined wind and wave loading on a monopile wind turbine. 

\subsubsection{Operational state}
\begin{figure}[h!]
\centering
\subfigure[]{\label{fig:Operational_bladeFn}\includegraphics[width=0.6\textwidth]{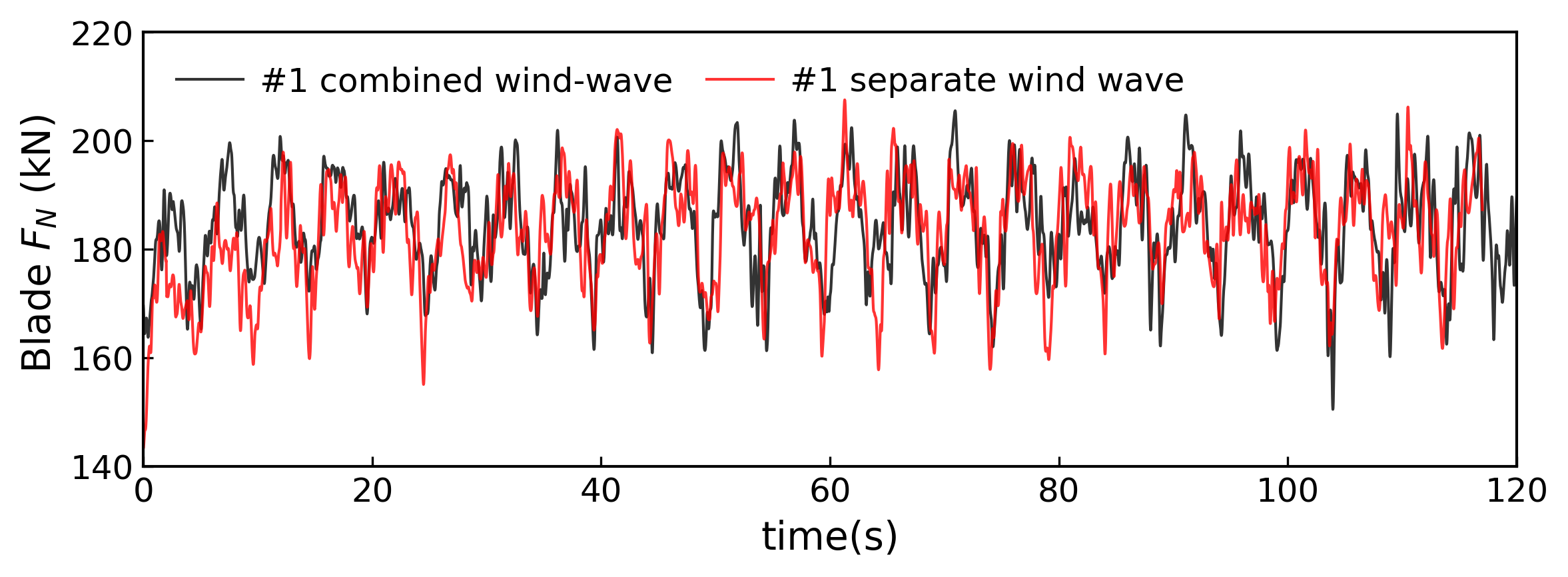}}
\subfigure[]{\label{fig:Operational_bladeFt}\includegraphics[width=0.6\textwidth]{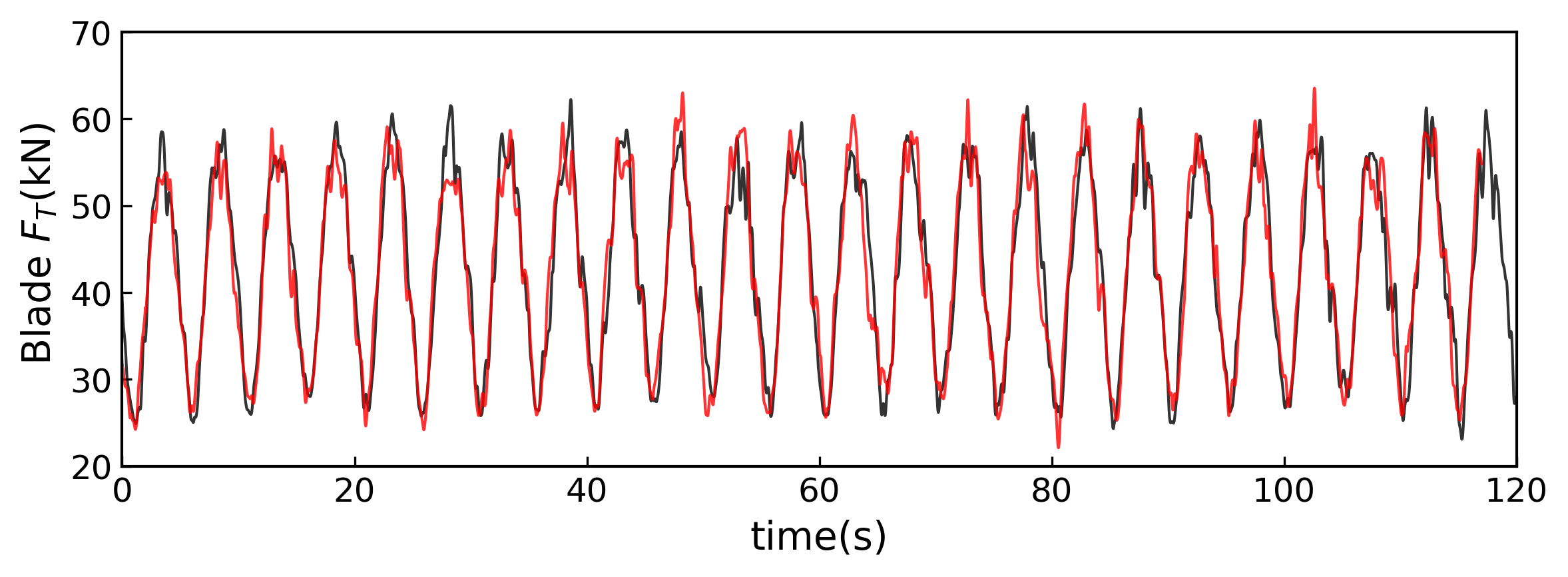}}
\caption{Time history of shear forces at the blade root under coupled and separated wind-wave fields during operational conditions. (a) Normal force; (b) Tangential force.}
\label{fig:Operational_bladeF}
\end{figure}

\begin{figure}[h!]

\centering\includegraphics[width=0.9\linewidth]{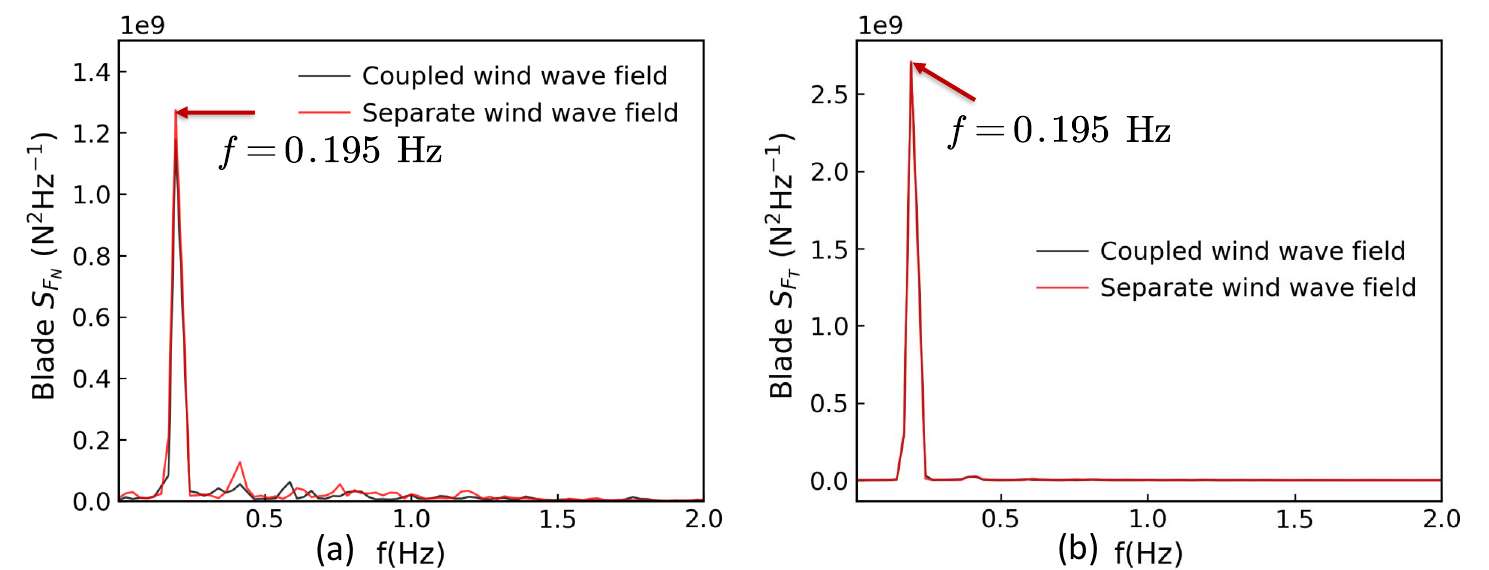}.
\caption{Spectrum of blade root shear force in the normal and tangential directions under coupled wind-wave fields and separated wind-wave fields during operational conditions. (a) Spectrum of normal force $S_{F_N}$; (b) Spectrum of tangential force $S_{F_T}$.}
\label{fig:OperationalBladeFPSD}
\end{figure}

In case 1, the wind speed at the hub height (87.6 m) is 12 m/s. The operational rotor speed is 12 mph. Fig. \ref{fig:Operational_bladeF} illustrates the time history of shear force at the blade root. For clarity, only one blade result is presented since the shear forces at the other two blades are similar. The black lines represent the shear force at the blade root under the coupled wind and wave fields, and the red lines illustrate the shear force under separate wind wave fields. Comparing the shear forces under coupled and separate wind and wave fields, it can be observed that the shear forces are close. This suggests that, under operational conditions, the wind-wave coupling effect on the blade shear force is minimal and can be disregarded. The lower normal force observed before $t=20$ s corresponds to the initial start-up phase of the wind turbine. Fig. \ref{fig:OperationalBladeFPSD} shows the power spectrum density of the normal force and the tangential force. The peak at a frequency equal to 0.195 Hz corresponds to the blade rotation frequency (0.2 Hz). Table \ref{table:BladeShearForcesOperation} presents the statistical values of the normal and tangential forces applied at the blade roots. During operational conditions, the average difference in shear force between coupled wind-wave fields and separate wind-wave fields is within $1\%$, while the maximum difference in shear force is within $4\%$. Additionally, the disparity in the standard deviation of the shear force under the two wind fields is within $3\%$. Table \ref{table:BladeShearForcesOperation} presents the statistical values of the bending moment experienced at the blade roots. The differences in the average, maximum, and standard deviation of bending moments at each blade between coupled and separate wind and wave fields are negligible. 

\begin{table}[h]
\centering
\caption{Shear forces at blade roots under operational conditions}
\label{table:BladeShearForcesOperation}
\begin{tabular}{l | l l l| l l l| l l l}
\hline
\multicolumn{1}{c|}{$F_N$} & \multicolumn{3}{c|}{Average (kN)} & \multicolumn{3}{c|}{Maximum (kN)} & \multicolumn{3}{c}{RMS (kN)}\\
\hline
{No.} &Coupled& sepa. &diff. &Coupled& sepa.&diff.&Coupled& sepa.&diff.\\
\hline
1 & 184.6  & 184.4 & 0.12$\%$ & 205.5 &207.5& -0.96$\%$ & 9.03 &9.27& -2.50$\%$\\
2 & 184.6  & 184.1 & 0.23$\%$ & 210.0 &203.1& 3.39$\%$ & 9.05 &9.17& -1.32$\%$\\
3 & 184.5  & 184.3 & 0.08$\%$ & 210.4 &208.1& 1.14$\%$ &9.06 & 9.07&-0.16$\%$\\
\hline
\multicolumn{1}{c|}{$F_T$} & \multicolumn{3}{c|}{Average (kN)} & \multicolumn{3}{c|}{Maximum (kN)} & \multicolumn{3}{c}{RMS (kN)}\\
\hline
1 & 43.02  & 43.07 & 0.12$\%$ &62.23 &63.53& -2.04$\%$ & 10.57 & 10.64&-0.70$\%$\\
2 & 43.10  & 43.03 & 0.16$\%$ & 62.46 &61.56& 1.46$\%$ & 10.59 & 10.59&0.01$\%$\\
3 & 42.89  & 42.87 & 0.03$\%$ & 63.48 &62.31& 1.88$\%$ & 10.72 & 10.62&0.92$\%$\\
\hline
\end{tabular}
\end{table}

\begin{table}[h]
\centering
\caption{Bending moment at blade roots under operational conditions}
\label{table:BladeBendingMomentOperation}
\begin{tabular}{l | l l l| l l l| l l l}
\hline
\multicolumn{1}{c|}{$M_{out}$} & \multicolumn{3}{c|}{Average (MNm)} & \multicolumn{3}{c|}{Maximum (MNm)} & \multicolumn{3}{c}{RMS (MNm)}\\
\hline
{No.} &Coupled& sepa.&diff. &Coupled& sepa.&diff.&Coupled& sepa.&diff.\\
\hline
1 & 7.50  & 7.49 & 0.18$\%$ & 8.36 &8.46& -1.18$\%$ & 0.365 &0.369& -1.14$\%$\\
2 & 7.50  & 7.48 & 0.43$\%$ & 8.54 &8.20& 4.10$\%$ & 0.358 &0.368& -2.95$\%$\\
3 & 7.49  & 7.48 & 0.17$\%$ & 8.55 &8.55& 0.00$\%$ &0.366 & 0.365&0.04$\%$\\
\hline
\multicolumn{1}{c|}{$M_{in}$} & \multicolumn{3}{c|}{Average (MNm)} & \multicolumn{3}{c|}{Maximum (MNm)} & \multicolumn{3}{c}{RMS (MNm)}\\
\hline
1 & 1.518  & 1.524 & -0.45$\%$ &2.278 &2.360& -3.43$\%$ & 0.429 & 0.432&-0.79$\%$\\
2 & 1.526  & 1.523 & 0.20$\%$ & 2.337 &2.245& 4.10$\%$ & 0.431 & 0.431&0.00$\%$\\
3 & 1.525  & 1.516 & 0.59$\%$ & 2.339 &2.317& 0.94$\%$ & 0.432 & 0.433&-0.33$\%$\\
\hline
\end{tabular}
\end{table}

\begin{figure}[h!]
\centering
\subfigure[]{\label{fig:Operational_TowerFx}\includegraphics[width=0.6\textwidth]{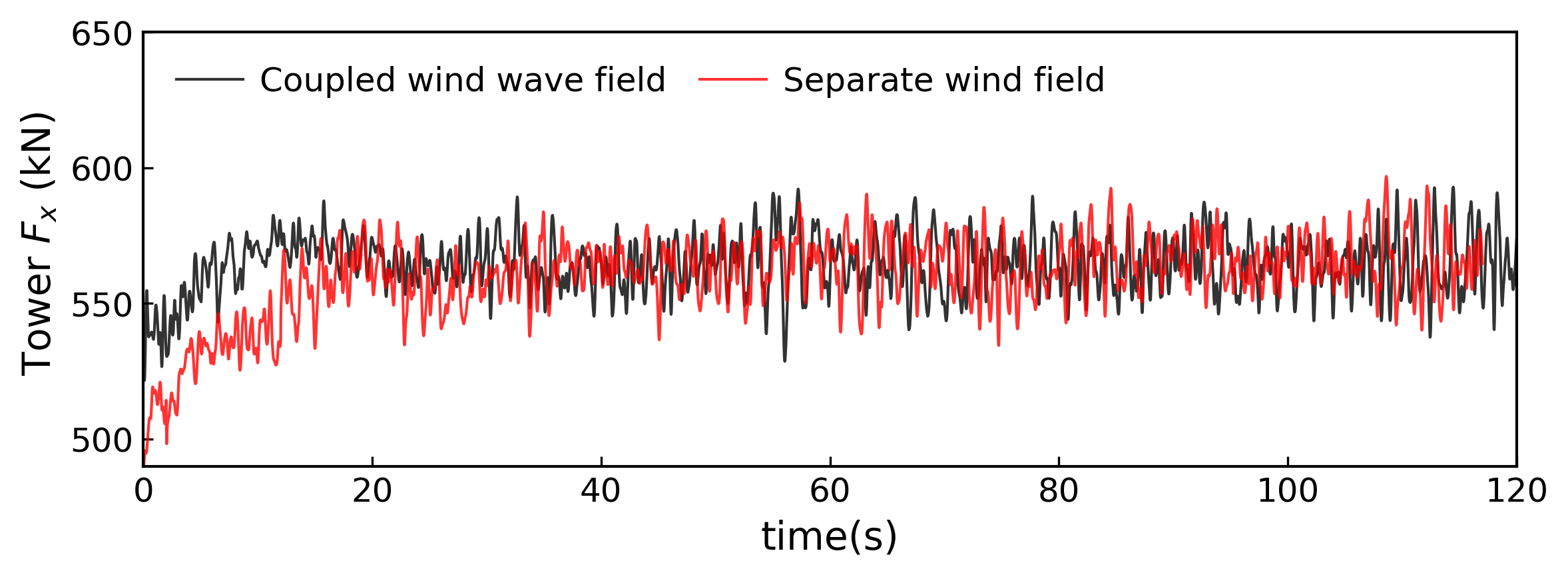}}
\subfigure[]{\label{fig:Operational_TowerFy}\includegraphics[width=0.6\textwidth]{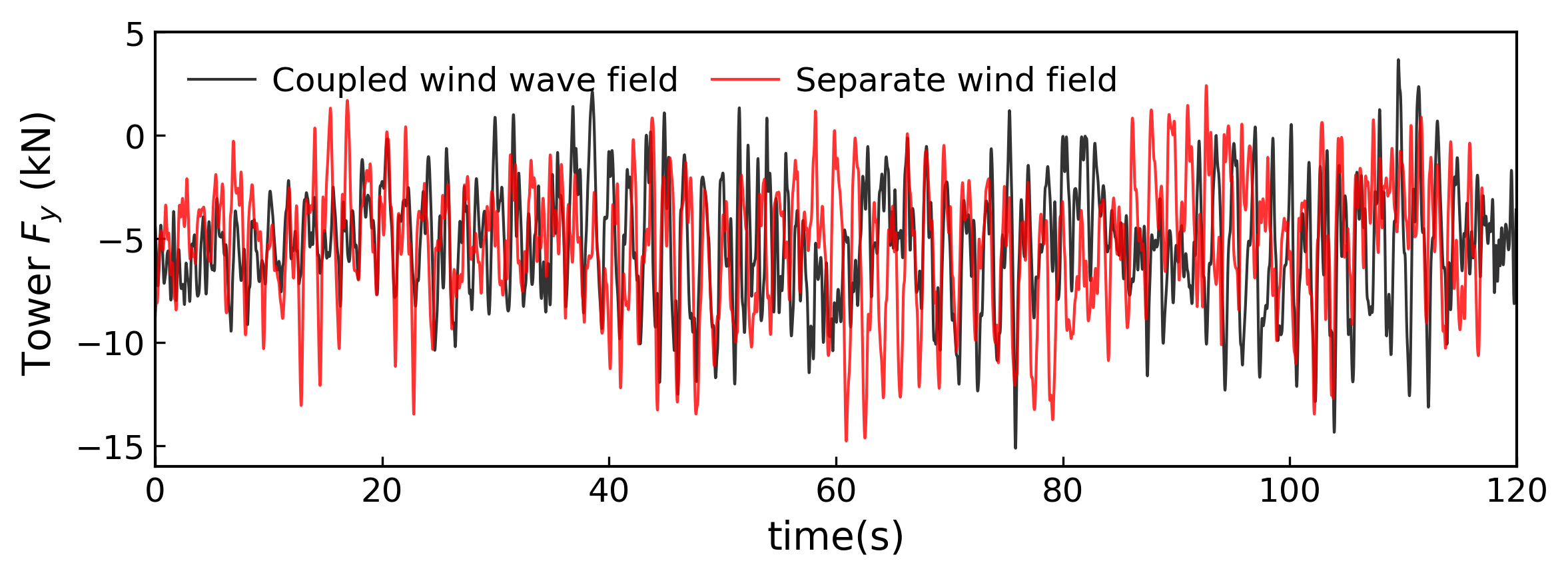}}
\caption{Time history of shear forces at the bottom of the tower during operational conditions. (a) Shear force in the $x$ direction ($F_x$); (b) Shear force in the $y$ direction ($F_y$).}
\label{fig:Operational_TowerF}
\end{figure}

\begin{figure}[h!]

\centering\includegraphics[width=0.9\linewidth]{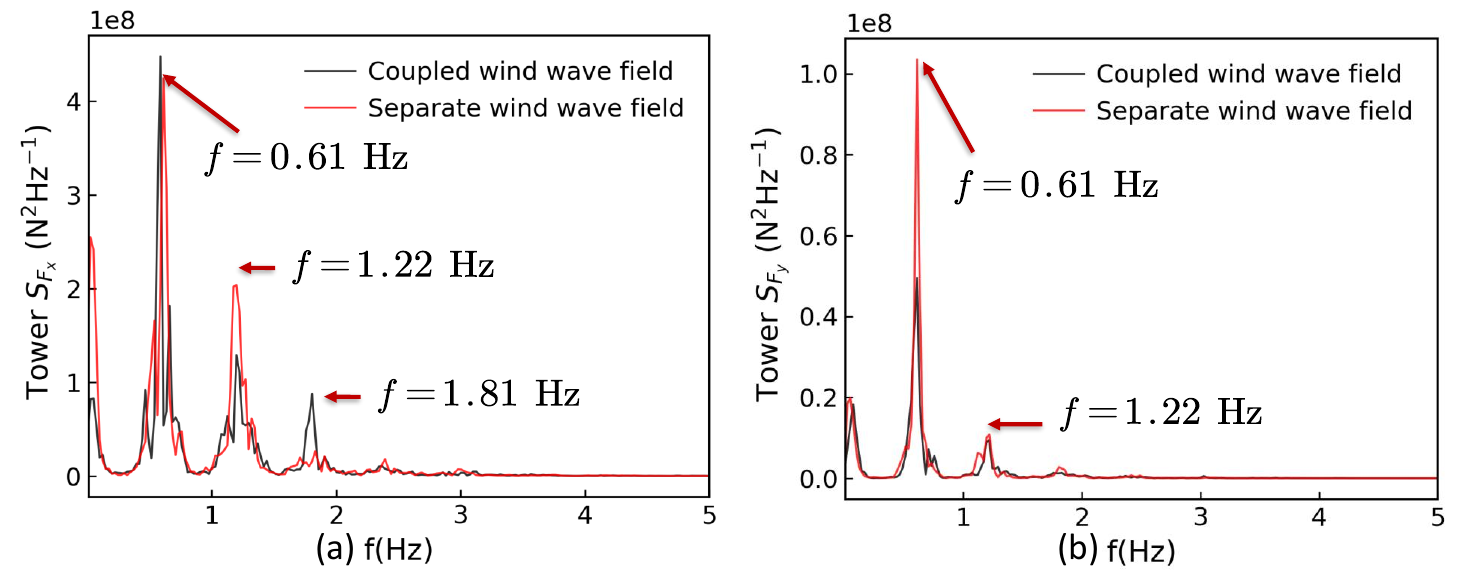}.
\caption{Spectrum of shear forces at the bottom of the tower in the normal and tangential directions under coupled wind-wave fields and separate wind-wave fields during operational conditions. (a) Spectrum of normal force $S_{F_N}$; (b) Spectrum of tangential force $S_{F_T}$.}
\label{fig:OperationaTowerFPSD}
\end{figure}

Fig. \ref{fig:Operational_TowerF} illustrates the shear force at the top of the transition piece, which corresponds to the bottom of the tower as shown in Fig. \ref{fig:MonopileOWT}. This position, denoted as position 2, is located above the wave peak. The forces acting on this position are primarily due to aerodynamic forces. Comparing the shear forces under coupled wind-wave fields and separate wind-wave fields at the tower bottom, it can be observed that they are similar. The power spectrum density of the shear force is depicted in Fig. \ref{fig:OperationaTowerFPSD}. A significant peak at a frequency of 0.61 Hz corresponds to the blade passing frequency 3P ($3\cdot0.2$ Hz). 

\begin{figure}[h!]
\centering
\subfigure[]{\label{fig:Operational_TowerFx}\includegraphics[width=0.6\textwidth]{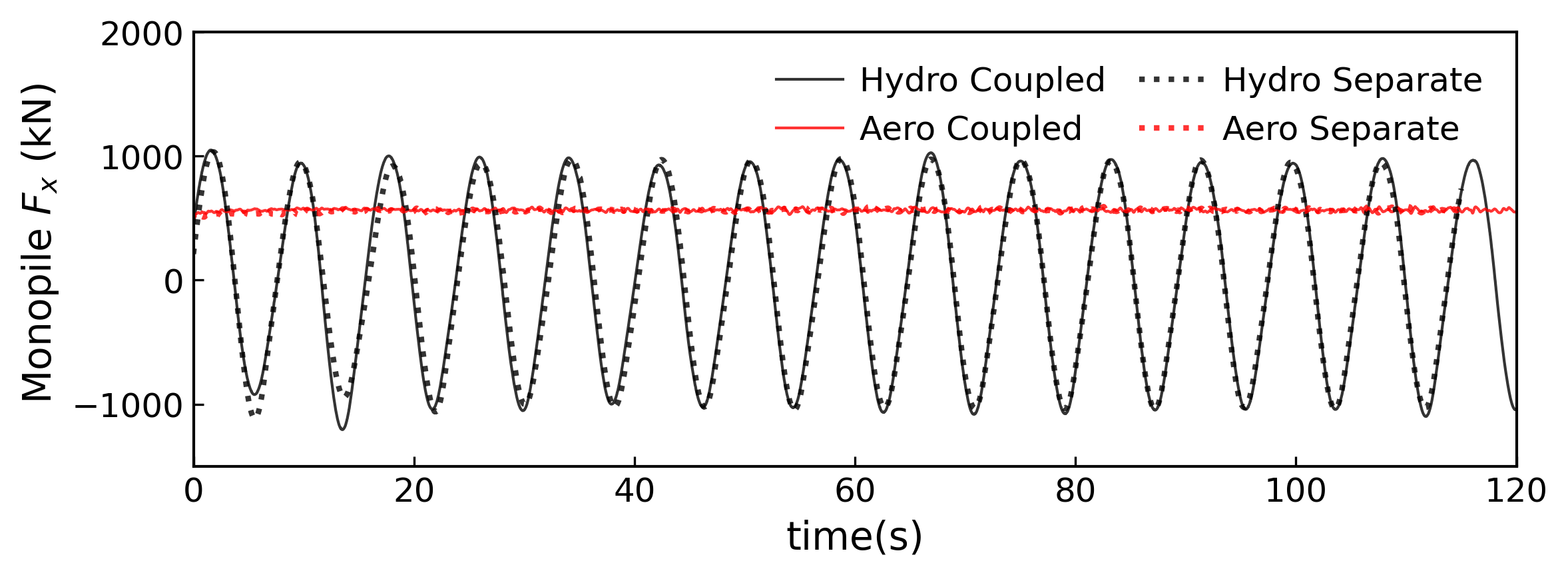}}
\subfigure[]{\label{fig:Operational_TowerFy}\includegraphics[width=0.6\textwidth]{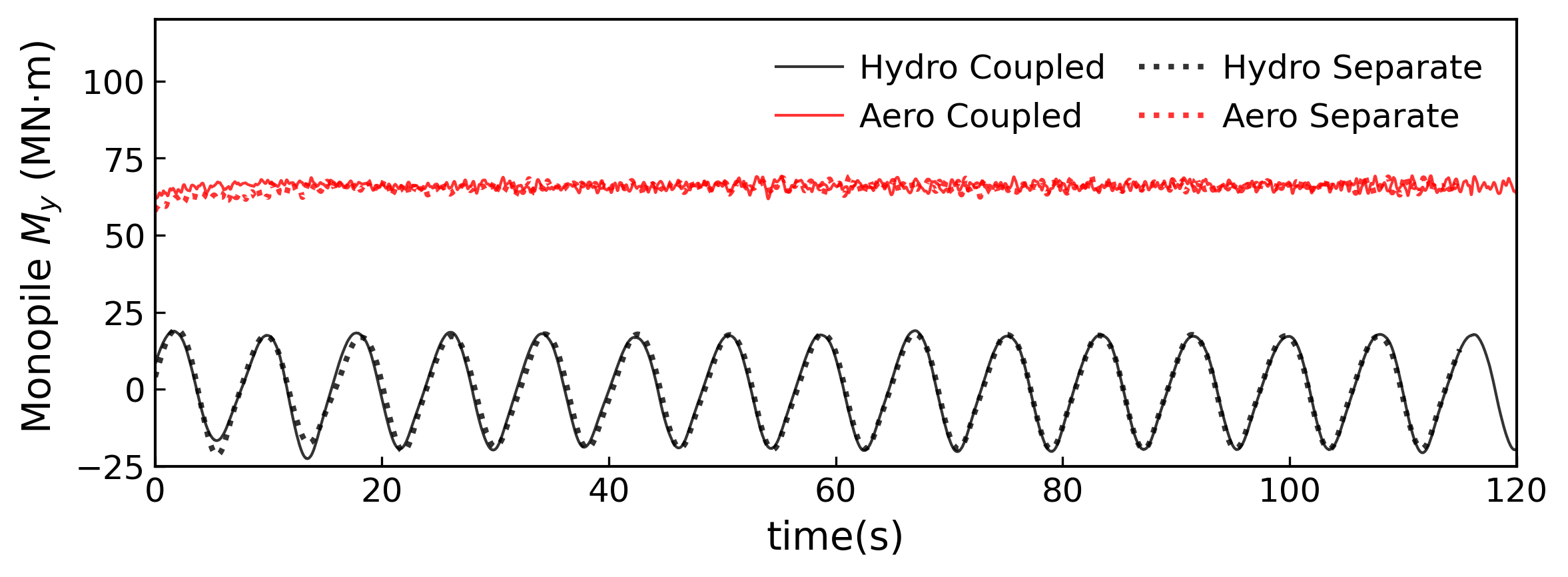}}
\caption{Time history of shear forces at the bottom of the monopile during operational conditions. (a) Shear force in the $x$ direction ($F_x$); (b) Shear force in the $y$ direction ($F_y$). The ``hydro coupled" and ``aero coupled" represent the hydrodynamic loading and aerodynamic loading under coupled wind and wave fields; The ``hydro separate" and ``aero separate" represent the hydrodynamic loading and aerodynamic loading under seperate wind and wave fields.}
\label{fig:Operational_MonopileF}
\end{figure}

Fig. \ref{fig:Operational_MonopileF} displays the shear force along the wind direction and the out-of-plane bending moment ($M_y$) at the bottom of the monopile. This corresponds to position 3 in Fig. \ref{fig:MonopileOWT}. The monopile is submerged underwater and subjected to both aerodynamic and hydrodynamic loadings. In Fig. \ref{fig:Operational_MonopileF}, the shear force induced by aerodynamic loading is represented by the red lines, while the shear force induced by hydrodynamic loading is depicted by the black lines. The average shear force is primarily caused by aerodynamic loading and the fluctuations of the shear force are mainly contributed by wave loading. Furthermore, the bending moment is mainly caused by aerodynamic loading, whereas hydrodynamic loading influences the variance of the bending moment. The loadings under coupled wind and wave fields are depicted by the solid lines, and the loadings under separate wind and wave fields are represented by the dashed lines. Upon comparing the loadings under the two wind-wave conditions, it is observed that the loadings acting at the bottom of the monopile are close.

Table \ref{table:SummaryOperation} summarizes the statistical loading applied to the nacelle, the bottom of the tower, and the bottom of the monopile. The variables are defined as follows: $M_t^{ro}$ represents the torque acting on the shaft, $F_x^{na}$ and $F_y^{na}$ indicate the forces acting at the nacelle in the along-wind and cross-wind directions, respectively. $F_x^{to}$ and $F_y^{to}$ denote the shear loading at the bottom of the tower, $M_x^{to}$ and $M_y^{to}$ represent the bending moments at the bottom of the tower, and $F_x^{mo}$ and $M_y^{mo}$ denote the shear force and bending moment at the bottom of the monopile. Comparing the loadings under combined and separated wind-wave fields, as listed in Table \ref{table:SummaryOperation}, the relative difference in average and maximum loading on the nacelle, the bottom of the tower, and the bottom of the monopile is within $3\%$, and the relative difference in standard deviation is within $9\%$. 

\begin{table}[h]
\centering
\caption{Summary of loadings during wind turbine operation}
\label{table:SummaryOperation}
\begin{tabular}{l | l l l| l l l| l l l}
\hline
\multicolumn{1}{c|}{} & \multicolumn{3}{c|}{Average} & \multicolumn{3}{c|}{Maximum} & \multicolumn{3}{c}{RMS}\\
\hline
{} &Coup.& sepa.&diff. &Coup.& sepa.&diff.&Coup.& sepa.&diff.\\
\hline
rotor $M_t^{ro}$ (MNm) & 4.154  & 4.151 & 0.07$\%$ & 4.522 &4.519& 0.07$\%$ &0.120 & 0.127&-1.50$\%$\\
nacelle $F_x^{na}$ (kN) & 554.0  & 552.8 & 0.21$\%$ & 582.1 &583.3& -0.29$\%$ & 9.5 &10.1& -5.4$\%$\\
nacelle $F_y^{na}$ (kN) & 4.78  & 4.67 & 2.20$\%$ & -13.97 &-14.04& -0.51$\%$ & 2.67 &2.91& 8.3$\%$\\

\hline
tower $F_x^{to}$ (kN) & 564.9  & 563.8 & 0.20$\%$ &592.9 &596.8& -0.66$\%$ & 9.66 & 10.11&-4.41$\%$\\
tower $F_y^{to}$ (kN) & -5.31  & -5.23 & 1.38$\%$ & -15.13 &-14.78& 2.34$\%$ & 2.81 & 3.06&-8.03$\%$\\
tower $M_x^{to}$ (MNm) & -0.346  & -0.344 & 0.74$\%$ & -1.031 &-1.051& -1.96$\%$ & 0.202 & 0.221&-8.27$\%$\\
tower $M_y^{to}$ (MNm) & 43.42  & 43.35 &0.19$\%$ & 45.60 &45.85& -0.56$\%$ & 0.764 &0.78&-4.81$\%$\\
\hline
monopile $F_x^{mo}$ (kN) & 518.7  & 531.0 & -2.31$\%$ &1606.5 &1545.1& 3.98$\%$ & 710.9 & 693.8&2.46$\%$\\
monopile $M_y^{mo}$ (MNm) & 64.88  & 64.79 & 0.14$\%$ & 87.28 &85.37& 2.24$\%$ & 13.13 & 12.83&2.26$\%$\\
\hline
\end{tabular}
\end{table}

\subsubsection{Cut-off state with pitch angle of 90$^{\circ}$}
\label{s:pitchAngleof90}
\begin{figure}[h!]

\centering\includegraphics[width=0.9\linewidth]{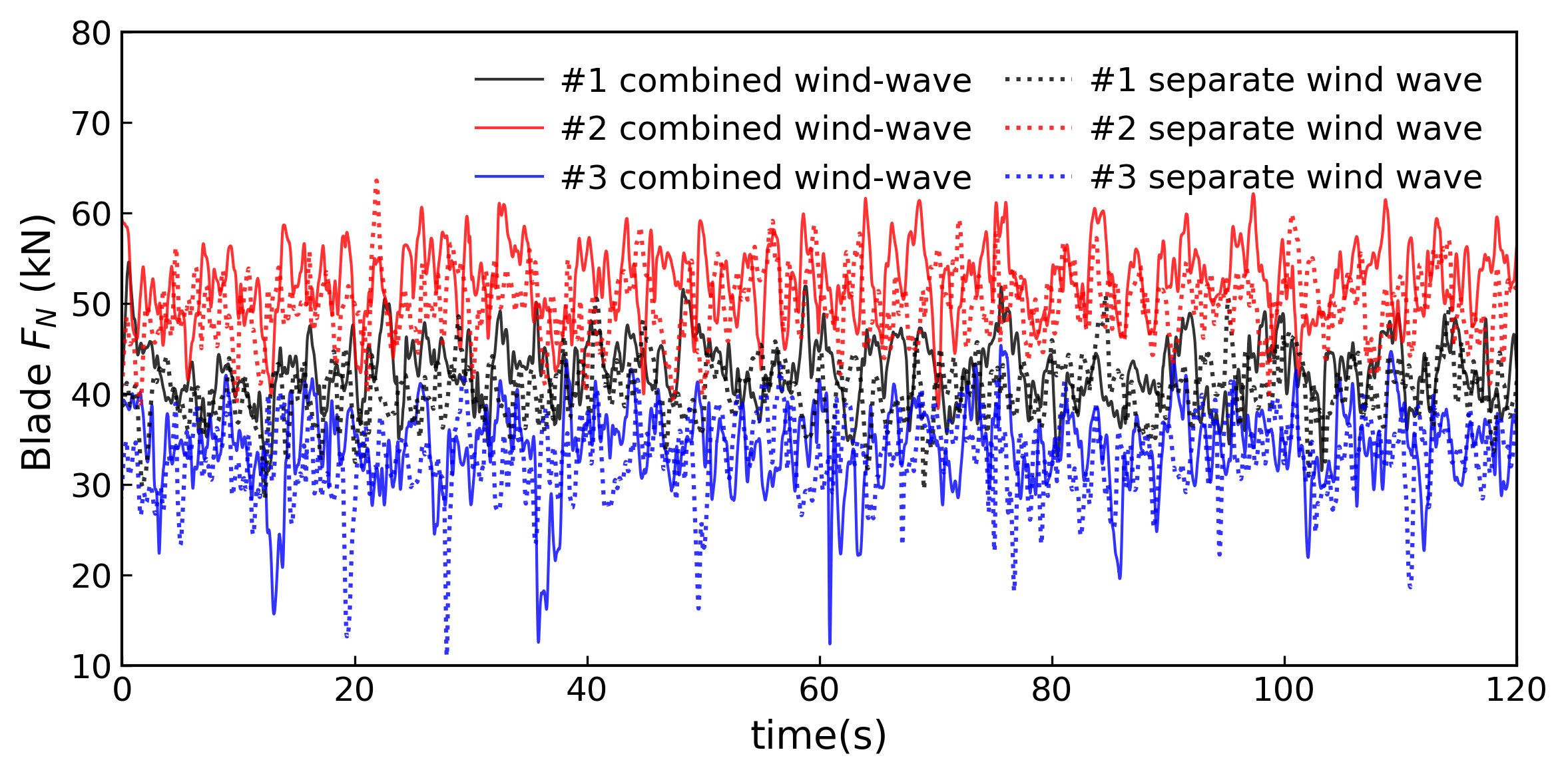}.
\caption{Normal force acting at blade roots under extreme conditions with a pitch angle of 90$^{\circ}$ }
\label{fig:Extreme_bladeFn}
\end{figure}

\begin{figure}[h!]

\centering\includegraphics[width=0.9\linewidth]{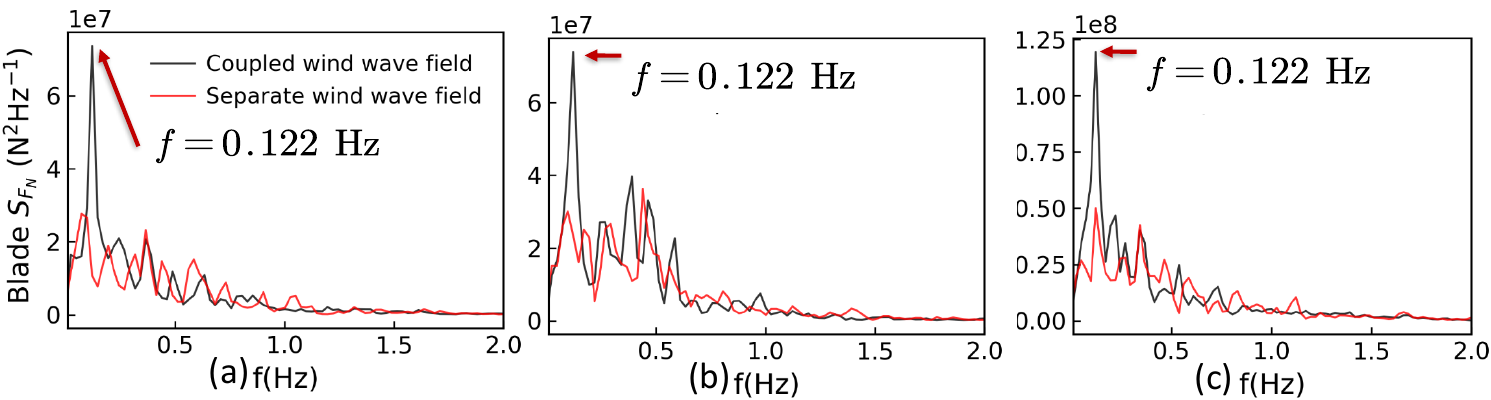}.
\caption{Spectrum of the normal force acting at blade roots under extreme conditions with a pitch angle of 90$^{\circ}$ }
\label{fig:Extreme_bladeFn_PSD}
\end{figure}

Under extreme wind conditions, where the wind speed exceeds the cut-off wind speed of the wind turbine, the turbine undergoes shutdown, and the blades are adjusted to align with the wind direction. In order to analyze the loading on wind turbines during these cut-off states under extreme wind and wave conditions, simulations are conducted for case 2 as outlined in Table \ref{table:WindTurbineCondition}. Fig. \ref{fig:Extreme_bladeFn} illustrates the normal force acting at the blade roots under extreme conditions with a pitch angle of 90$^{\circ}$. The variation in forces at the three blade roots is primarily attributed to differences in blade location and wind attack angles. When comparing the normal force under coupled wind and wave fields with that under separate wind and wave fields, it becomes evident that the blades experience more significant fluctuations in normal force under coupled wind and wave fields. The spectrum of the normal force at the three blades is depicted in Fig \ref{fig:Extreme_bladeFn_PSD}. Under coupled wind and wave fields, a peak is observed at a frequency of 0.122 Hz, corresponding to the wave frequency. The presence of extremely high waves (H=10 m) intensified the wind turbulence, thereby amplifying the fluctuations of aerodynamic loading on the blades. Table \ref{table:ExtremeBladeShearForces} presents the shear force at the blade roots. When comparing the loading under coupled and separated wind wave fields, the wind-wave coupling effect leads to an increase in the average shear force at the blade root. Specifically, the average normal force experiences an increase of up to $6.9\%$, and the average tangential force undergoes an increase of up to $10.4\%$. High waves induce wind turbulence, resulting in significant fluctuations in wind loadings. This is reflected in the standard deviation of the shear force, with the standard deviation of the normal force increasing by up to $18.2\%$ in blade 3. Additionally, Table \ref{table:ExtremeBladeBendingMoment} lists the bending moment at the blade root. The coupling effect between wind and waves amplifies the fluctuation of loading, as evidenced by the increase in the standard deviation of the out-of-plane moment, which increases by $25.4\%$ in blade 3. Comparing the shear force and bending moment during wind turbine shutdown conditions (in Table \ref{table:ExtremeBladeShearForces} and Table \ref{table:ExtremeBladeBendingMoment}) with those during operational conditions (in Table \ref{table:BladeShearForcesOperation} and Table \ref{table:BladeBendingMomentOperation}), some notable differences can be observed. During wind turbine shutdown conditions, the blades experience smaller normal forces and out-of-plane moments, indicating a reduced overall load on the blades. However, they exhibit larger tangential forces and in-plane moments, suggesting higher rotational forces exerted on the blades. 

\begin{table}[h]
\centering
\caption{Shear forces at blade roots under extreme conditions with a pitch angle of $90^{\circ}$}
\label{table:ExtremeBladeShearForces}
\begin{tabular}{l | l l l| l l l| l l l}
\hline
\multicolumn{1}{c|}{$F_N$} & \multicolumn{3}{c|}{Average (kN)} & \multicolumn{3}{c|}{Maximum (kN)} & \multicolumn{3}{c}{RMS (kN)}\\
\hline
{No.} &Coupled& sepa.&diff. &Coupled& sepa.&diff.&Coupled& sepa.&diff.\\
\hline
1 & 42.45  & 39.70 & 6.92$\%$ & 54.58 &50.92& 7.20$\%$ & 3.54&3.28& 7.87$\%$\\
2 & 52.30  & 48.95 & 6.84$\%$ & 62.14 &63.57& -2.26$\%$ & 4.03 &3.78& 6.47$\%$\\
3 & 34.05  & 32.61 & 4.41$\%$ & 45.38 &43.05& 5.41$\%$ &4.64 & 3.93&18.19$\%$\\
\hline
\multicolumn{1}{c|}{$F_T$} & \multicolumn{3}{c|}{Average (kN)} & \multicolumn{3}{c|}{Maximum (kN)} & \multicolumn{3}{c}{RMS (kN)}\\
\hline
1 & 142.5  & 135.8 & 4.92$\%$ &225.3 &236.7& 4.85$\%$ & 30.17 & 27.44&9.96$\%$\\
2 & 217.2  & 196.8 & 10.39$\%$ & 285.7 &283.2& 0.90$\%$ & 29.77& 26.87&10.83$\%$\\
3 & 128.5  & 121.2 & 6.08$\%$ & 221.8 &220.4& 0.65$\%$ & 33.74 & 31.15&8.31$\%$\\
\hline
\end{tabular}
\end{table}

\begin{table}[h]
\centering
\caption{Bending moment at blade roots under extreme conditions with a pitch angle of $90^{\circ}$}
\label{table:ExtremeBladeBendingMoment}
\begin{tabular}{l | l l l| l l l| l l l}
\hline
\multicolumn{1}{c|}{$M_{out}$} & \multicolumn{3}{c|}{Average (MNm)} & \multicolumn{3}{c|}{Maximum (MNm)} & \multicolumn{3}{c}{RMS (MNm)}\\
\hline
{No.} &Coupled& sepa.&diff. &Coupled& sepa.&diff.&Coupled& sepa.&diff.\\
\hline
1 & 0.357  & 0.342 & 4.02$\%$ & 0.517 &0.467& 10.59$\%$ & 0.059&0.051& 16.86$\%$\\
2 & 0.578 & 0.558& 3.57$\%$ & 0.743 &0.743& 0.00$\%$ & 0.094 &0.075& 24.16$\%$\\
3 & 0.184 & 0.190& -2.82$\%$ & 0.383 &0.369& 3.65$\%$ &0.137 & 0.110&25.37$\%$\\
\hline
\multicolumn{1}{c|}{$M_{in}$} & \multicolumn{3}{c|}{Average (MNm)} & \multicolumn{3}{c|}{Maximum (MNm)} & \multicolumn{3}{c}{RMS (MNm)}\\
\hline
1 & 3.698 & 3.518& 5.09$\%$ &6.431&7.084& -9.21$\%$ & 0.927 & 0.862&7.51$\%$\\
2 & 5.902  &5.244  & 12.54$\%$ & 8.597 &7.864& 9.32$\%$ & 1.051 & 0.927&13.39$\%$\\
3 & 4.252  &3.945 & 7.78$\%$ & 7.553&7.482& 0.94$\%$ & 1.177& 1.058&11.27$\%$\\
\hline
\end{tabular}
\end{table}

\begin{figure}[h!]
\centering
\subfigure[]{\label{fig:Extreme_TowerFx}\includegraphics[width=0.8\textwidth]{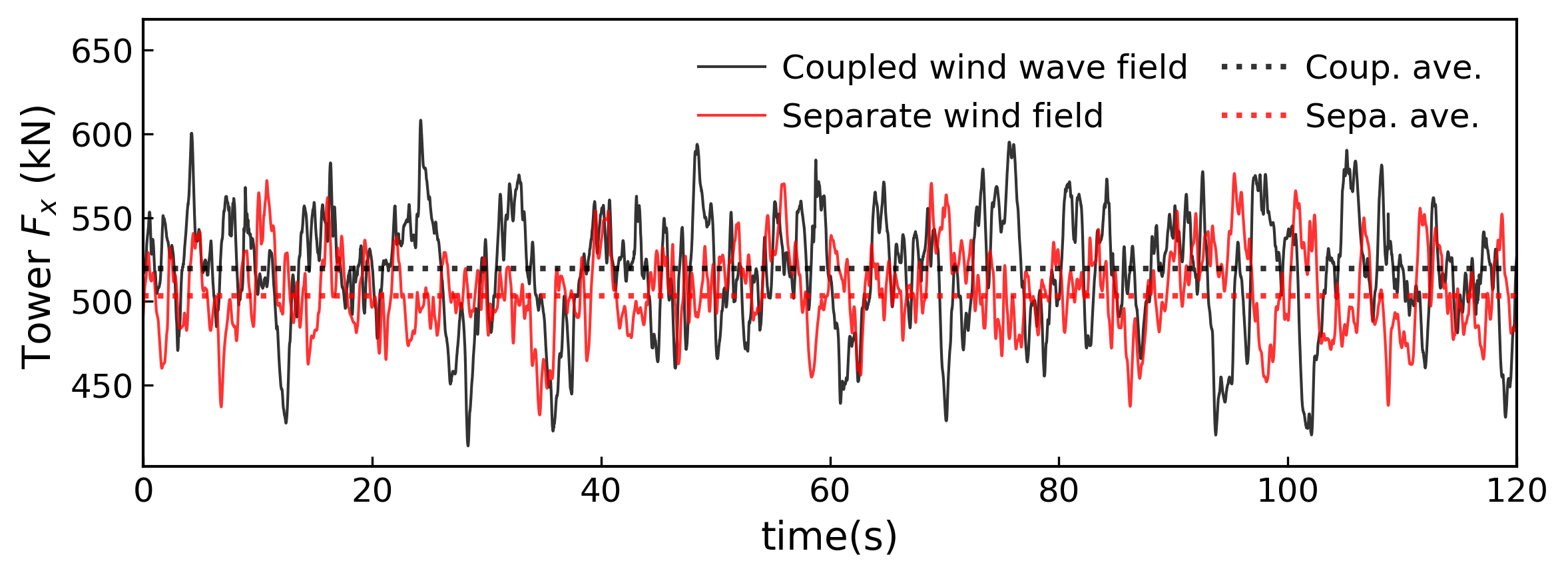}}
\subfigure[]{\label{fig:Extreme_TowerFy}\includegraphics[width=0.8\textwidth]{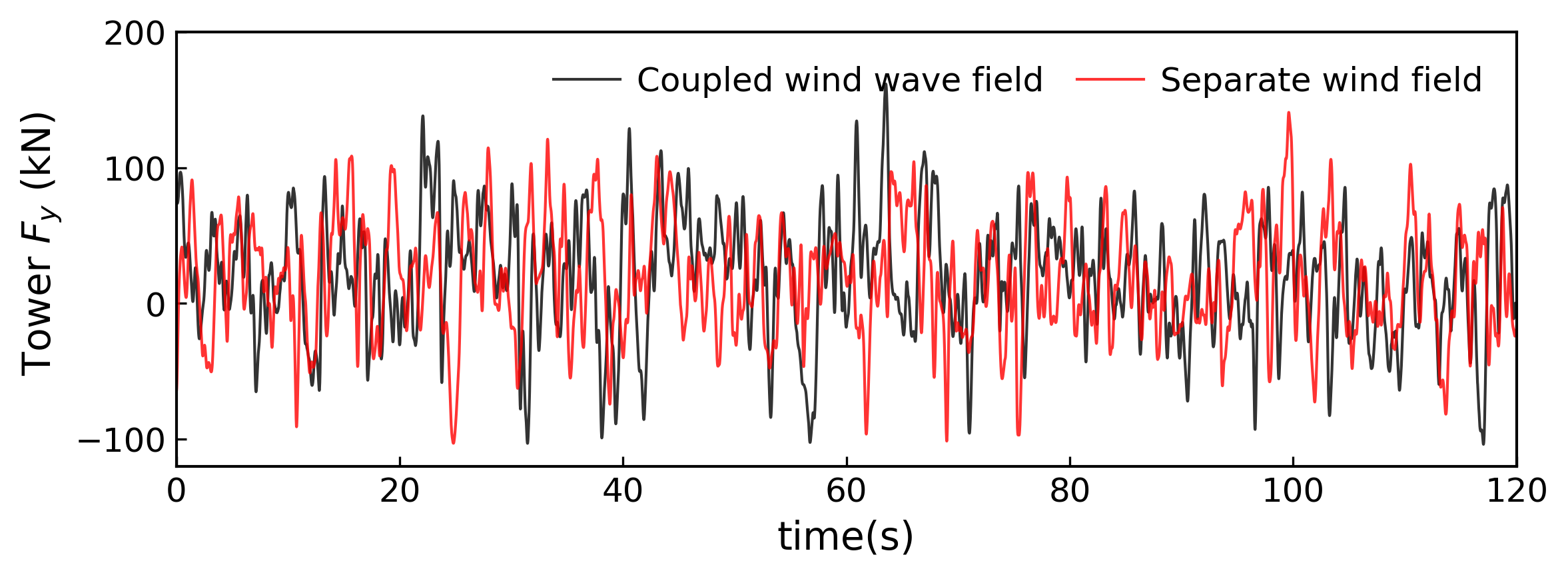}}
\caption{Time history of shear forces at the bottom of the tower during extreme conditions with the pitch angle of $90^{\circ}$. (a) Shear force in the $x$ direction ($F_x$); (b) Shear force in the $y$ direction ($F_y$).}
\label{fig:Extreme_TowerF}
\end{figure}

\begin{figure}[h!]
\centering
\subfigure[]{\label{fig:Extreme_TowerFx_PSD}\includegraphics[width=0.4\textwidth]{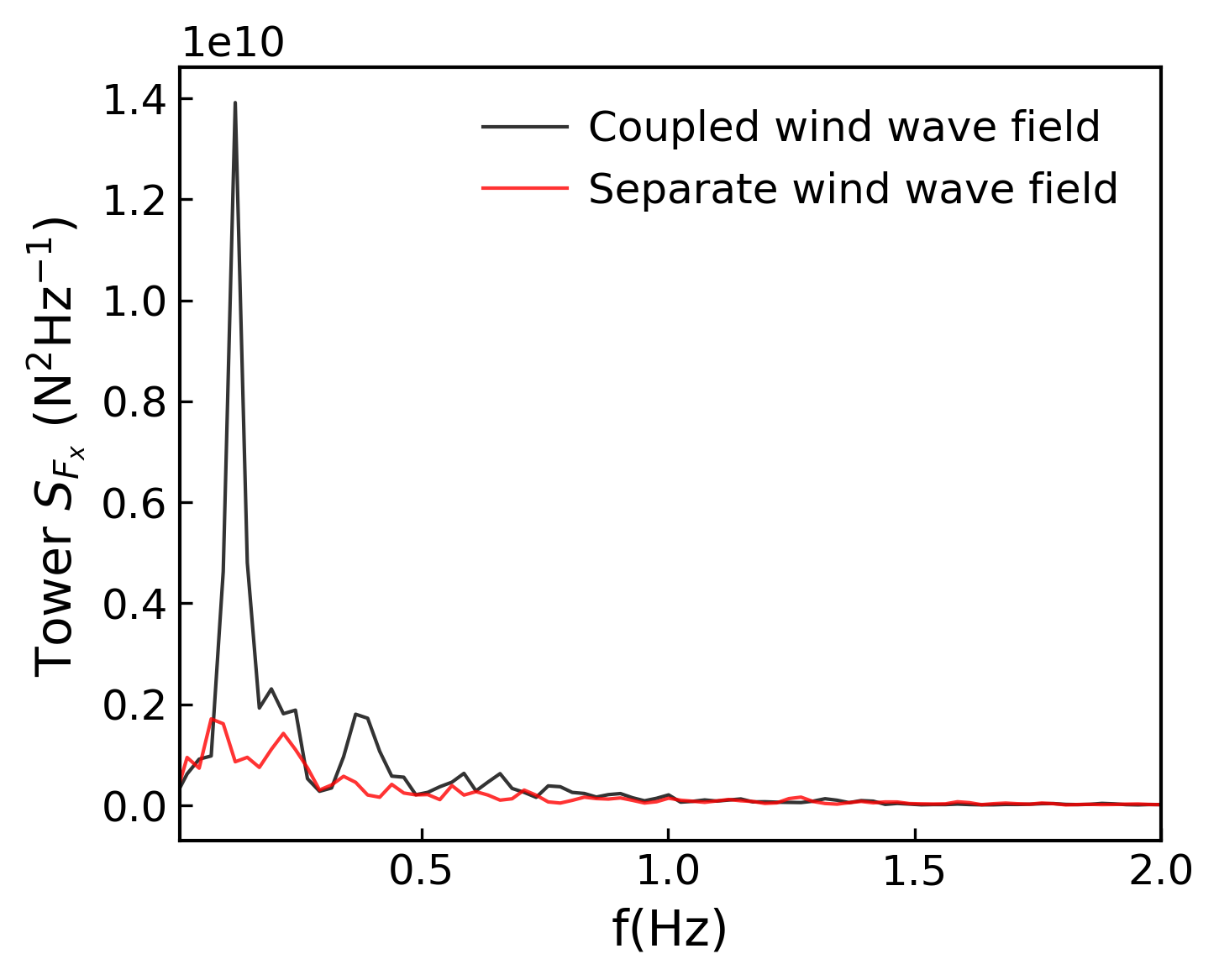}}
\subfigure[]{\label{fig:Extreme_TowerFy_PSD}\includegraphics[width=0.4\textwidth]{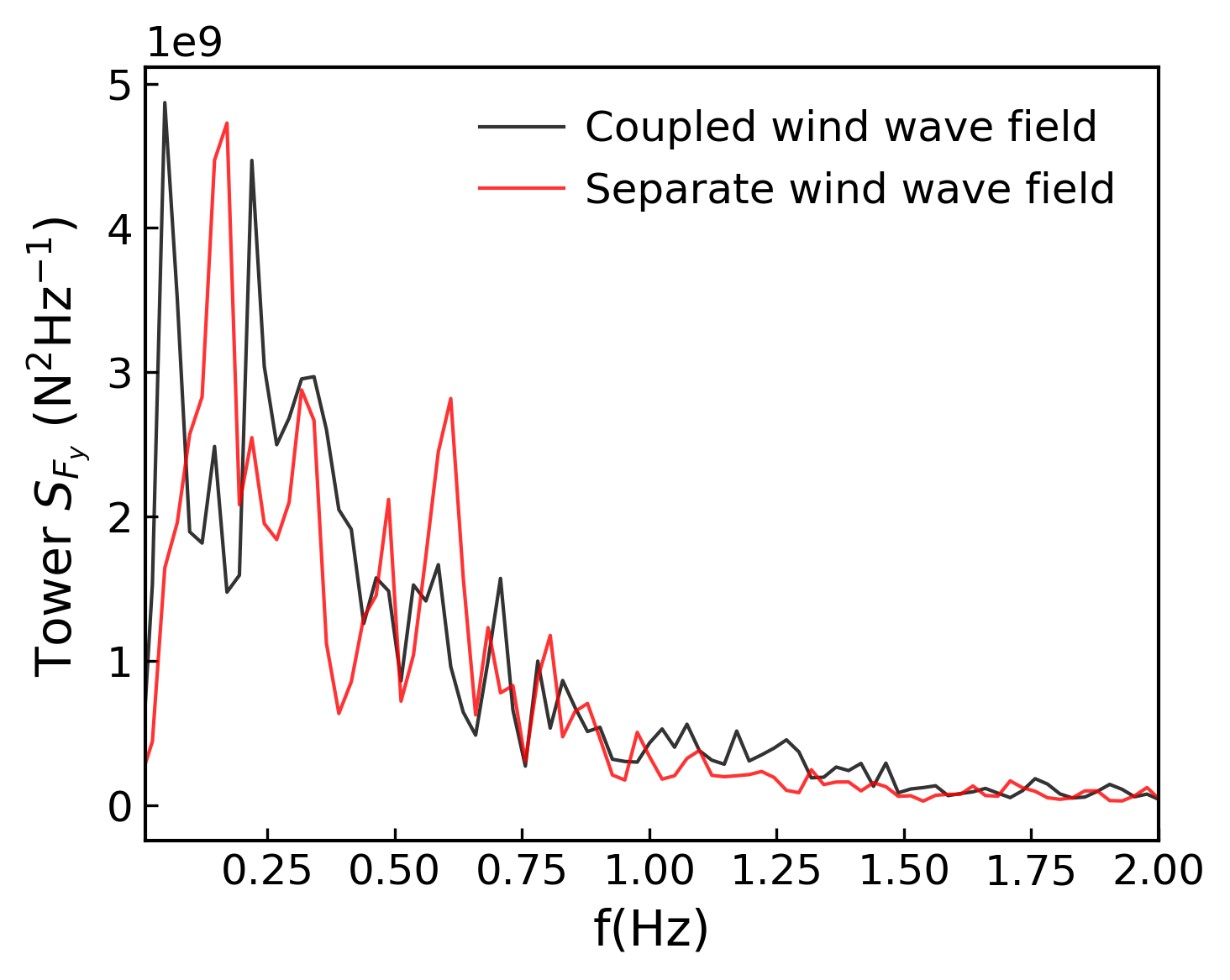}}
\caption{Spectrum of tower shear force under coupled wind wave fields and separate wind wave fields during extreme conditions with the pitch angle of $90^{\circ}$. (a) The spectrum of shear force in $x$ direction; (b)The spectrum of shear force in $y$ direction}
\label{fig:Extreme_TowerF_PSD}

\end{figure}

\begin{figure}[h!]
\centering
\subfigure[]{\label{fig:Extreme_TowerMx}\includegraphics[width=0.8\textwidth]{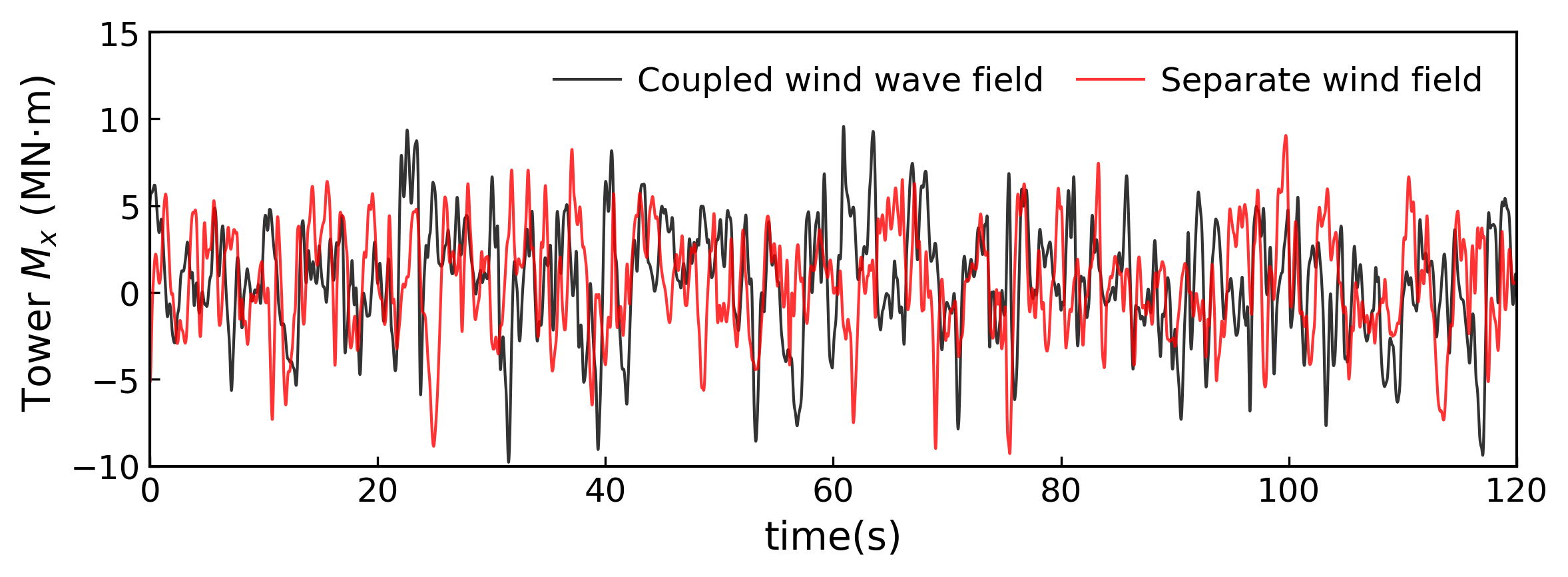}}
\subfigure[]{\label{fig:Extreme_TowerMy}\includegraphics[width=0.8\textwidth]{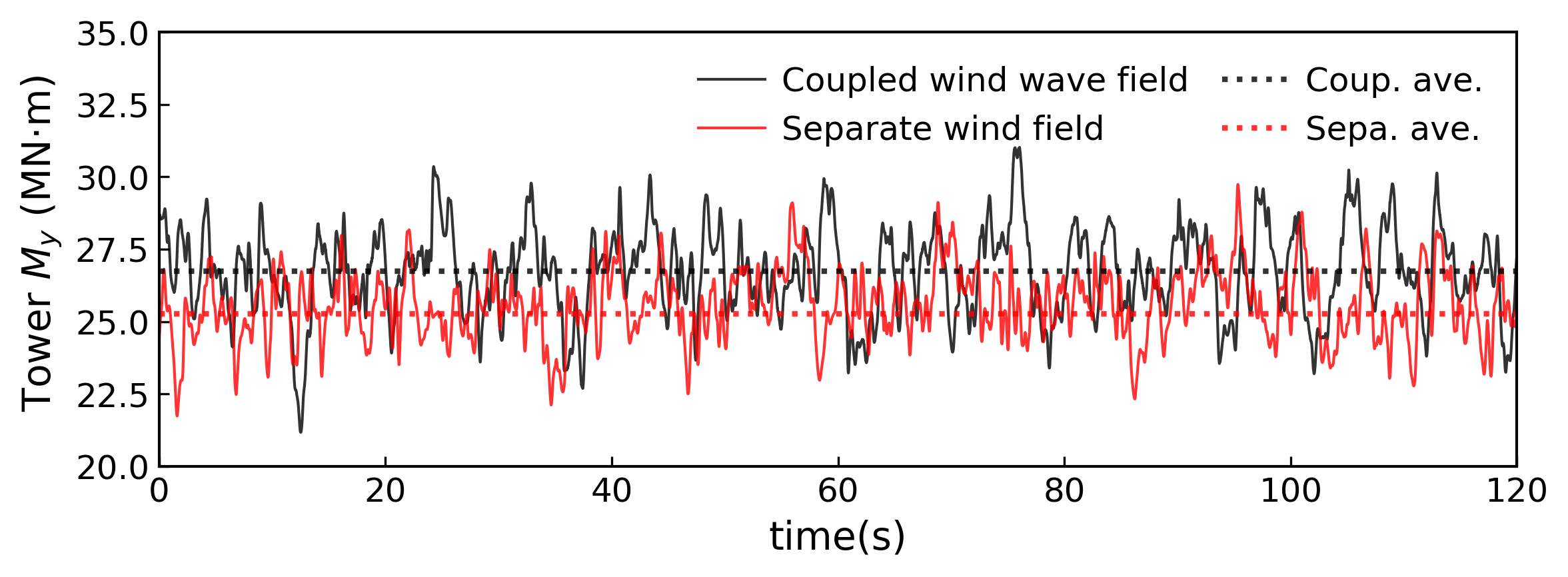}}
\caption{Time history of bending moment at the bottom of the tower during extreme conditions with the pitch angle of $90^{\circ}$. (a) Bending moment in the $x$ direction ($M_x$); (b) Bending moment in the $y$ direction ($M_y$)}
\label{fig:Extreme_TowerM}
\end{figure}

Fig. \ref{fig:Extreme_TowerF} displays the time history of the shear force at the bottom of the tower. The black lines represent the shear force induced by the coupled wind and wave fields, while the red lines illustrate the shear force under separated wind and wave fields. The dotted lines correspond to the average values of the shear force. As shown in Fig. \ref{fig:Extreme_TowerFx}, under the coupled wind and wave fields, the $F_x$ exhibits a larger average value compared to the separate wind and wave fields, with the average $F_x$ increased by $3.24\%$. Moreover, the variance of shear force is greater under the coupled wind and wave fields. The relative differences in the standard deviation of $F_x$ and $F_y$ between the coupled and separate wind and wave fields are up to $44.6\%$ and $13.1\%$, respectively. The spectrum of the shear force at the bottom of the tower is depicted in Fig. \ref{fig:Extreme_TowerF_PSD}. In particular, the spectrum of the shear force in the $X$ direction exhibits a noticeable peak at a frequency of 0.122 Hz under coupled wind and wave fields, as illustrated in Fig. \ref{fig:Extreme_TowerFx_PSD}. Conversely, the peak in the spectrum of the shear force in the $Y$ direction is not as prominent. As discussed in Section \ref{s:windEvolutionAlongWave}, the presence of waves induces turbulence primarily in the along-wind direction, thereby exerting a more significant impact on the wind loadings in the $X$ direction. Similarly, the bending moment in $y$ direction exhibits a larger average value under the coupled wind and wave fields compared to that under separated wind and wave fields, as shown in Fig. \ref{fig:Extreme_TowerMy}. The average $M_y$ is increased by $5.3\%$. Moreover, the variance of bending moment is greater under the coupled wind and wave fields. The relative differences in the standard deviation of $M_x$ and $M_y$ between the coupled and separate wind and wave fields are up to $16\%$ and $33\%$, respectively.

Table \ref{table:ExtremeSummary} presents a summary of the loadings acting on the wind turbine under extreme conditions, with a blade pitch angle of 90$^{\circ}$. It is observed that the coupled wind and wave effect leads to an increase in the average loading values, along with larger variations in the loading. Specifically, the maximum bending moment at the monopile bottom is increased by $7.0\%$ due to the coupled wind and wave effect. Comparing the loadings in Table \ref{table:ExtremeSummary} with those in Table \ref{table:SummaryOperation}, it can be observed that the $F_x^{to}$, $M_y^{to}$, and $M_y^{mo}$ under extreme wind and wave conditions have smaller values than those under operational conditions. This is primarily due to the successful control of the blade pitch angle, which allows the wind turbine blades to align with the wind direction and reduce the aerodynamic loading on the structure. By effectively adapting to extreme wind wave conditions, the offshore wind turbine can sustain challenging environmental conditions.

\begin{table}[h]
\centering
\caption{Summary of the loading under extreme conditions with blade pitch angle of 90$^{\circ}$}
\label{table:ExtremeSummary}
\begin{tabular}{l | l l l| l l l| l l l}
\hline
\multicolumn{1}{c|}{} & \multicolumn{3}{c|}{Average} & \multicolumn{3}{c|}{Maximum} & \multicolumn{3}{c}{RMS}\\
\hline
{} &Coup.& sepa.&diff. &Coup.& sepa.&diff.&Coup.& sepa.&diff.\\
\hline
rotor $M_t^{ro}$ (MNm) & -5.44  & -5.24 & 3.67$\%$ & -11.95&-13.01& -8.77$\%$ &2.21 & 2.28&-2.76$\%$\\
nacelle $F_x^{na}$ (kN) & 128.8  & 121.3 & 6.21$\%$ & 155.7&144.3& 7.89$\%$ & 9.019 &8.028& 12.35$\%$\\
nacelle $F_y^{na}$ (kN) &  13.46 & 12.67& 6.28$\%$ & 138.4&119.1& 16.15$\%$ & 41.47 &35.98& 15.28$\%$\\

\hline
tower $F_x^{to}$ (kN) & 519.9  & 503.6 & 3.24$\%$ &613.1 &576.3& 6.38$\%$ & 34.94 & 24.16&44.61$\%$\\
tower $F_y^{to}$ (kN) & 19.02  & 19.45 & -2.21$\%$ & 161.8&140.8& 14.86$\%$ &  42.77& 37.81&13.11$\%$\\
tower $M_x^{to}$ (MNm) & 0.674&0.707 &-4.72$\%$ & 9.565 &9.046& 5.73$\%$ & 3.246 & 2.798&16.05$\%$\\
tower $M_y^{to}$ (MNm) & 26.74  & 25.40 & 5.27$\%$ & 31.02 &29.73& 4.31$\%$ & 1.53 & 1.15&32.96$\%$\\
\hline
monopile $F_x^{mo}$ (kN) &  656.7 & 640.4 & 2.55$\%$ &3518.9 &3402.6& 3.42$\%$ & 1831.0 & 1788.4&2.38$\%$\\
monopile $M_y^{mo}$ (MNm) & 55.28  & 52.86& 4.58$\%$ & 117.2 &109.5&7.04$\%$ & 35.03 & 33.38&4.94$\%$\\
\hline
\end{tabular}
\end{table}

\subsubsection{Cut-off state with pitch angle of 60$^{\circ}$}
\begin{figure}[h!]

\centering\includegraphics[width=0.9\linewidth]{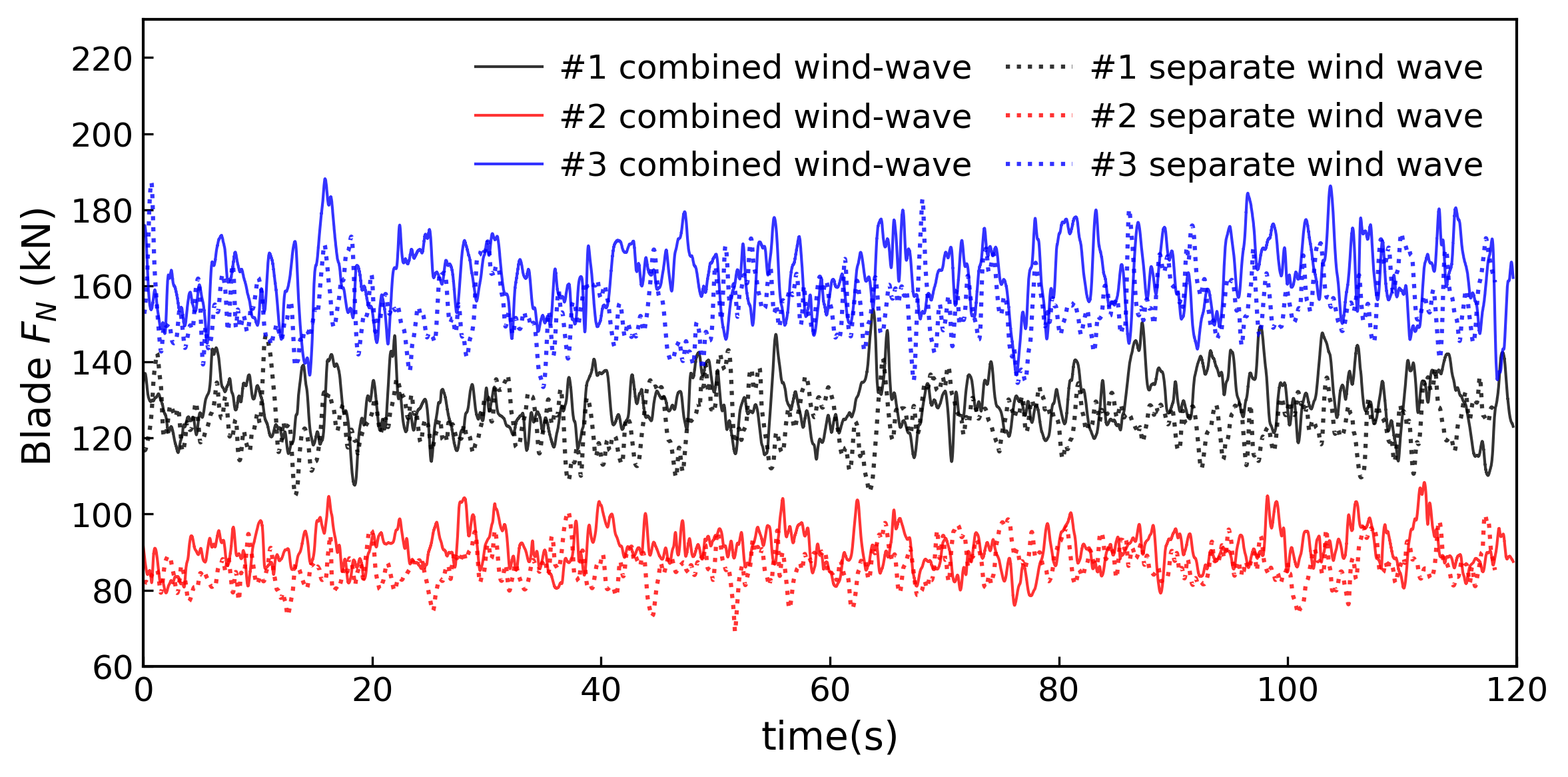}.
\caption{Normal force acting at blade roots under extreme conditions with a pitch angle of $60^{\circ}$}
\label{fig:ExtremePitch_blade}
\end{figure}

Under extreme circumstances, the blade pitch control may malfunction due to power supply failure. This section investigates the behavior of a wind turbine by simulating a fixed pitch angle of $60^{\circ}$. The aim is to examine how the turbine performs under challenging conditions without successful pitch control. Fig. \ref{fig:ExtremePitch_blade} displays the normal force acting on the three blades under both coupled wind-wave fields (solid lines) and separated wind and wave fields (dotted lines). It is evident from Fig. \ref{fig:ExtremePitch_blade} that the normal forces on the blades are higher when subjected to coupled wind and wave fields compared to separate wind and wave fields. This is consistent with the finding in section \ref{s:pitchAngleof90}. Table \ref{table:ExtremePitch_NormalForces} presents the maximum and average normal forces for each of the three blades. When comparing the normal force in coupled wind-wave fields to separated wind and wave fields, Blade 2 demonstrates the most significant increase in maximum normal force at the root, approximately 7.7 kN, resulting in a $7.7\%$ rise. The presence of waves induces wind turbulence near the wave surface, leading to higher peak loading on structures. 
Table \ref{table:ExtremePitch_BladeBendingMoment} presents the bending moment at the blade roots under extreme wind wave conditions. Comparing the loading with a pitch angle of $60^{\circ}$ in Table \ref{table:ExtremePitch_NormalForces} and Table \ref{table:ExtremePitch_BladeBendingMoment} to the loading with a pitch angle of $90^{\circ}$ in Table \ref{table:ExtremeBladeShearForces} and Table \ref{table:ExtremeBladeBendingMoment}, significant increases in shear force and bending moment at the blade roots can be observed due to the failed control of the blade pitch. The inability to properly adjust the blade pitch angle results in higher loads on the blades, leading to increased shear force and bending moment.

\begin{figure}[h!]
\centering
\subfigure[]{\label{fig:Axialforce_frequency_37}\includegraphics[width=0.8\textwidth]{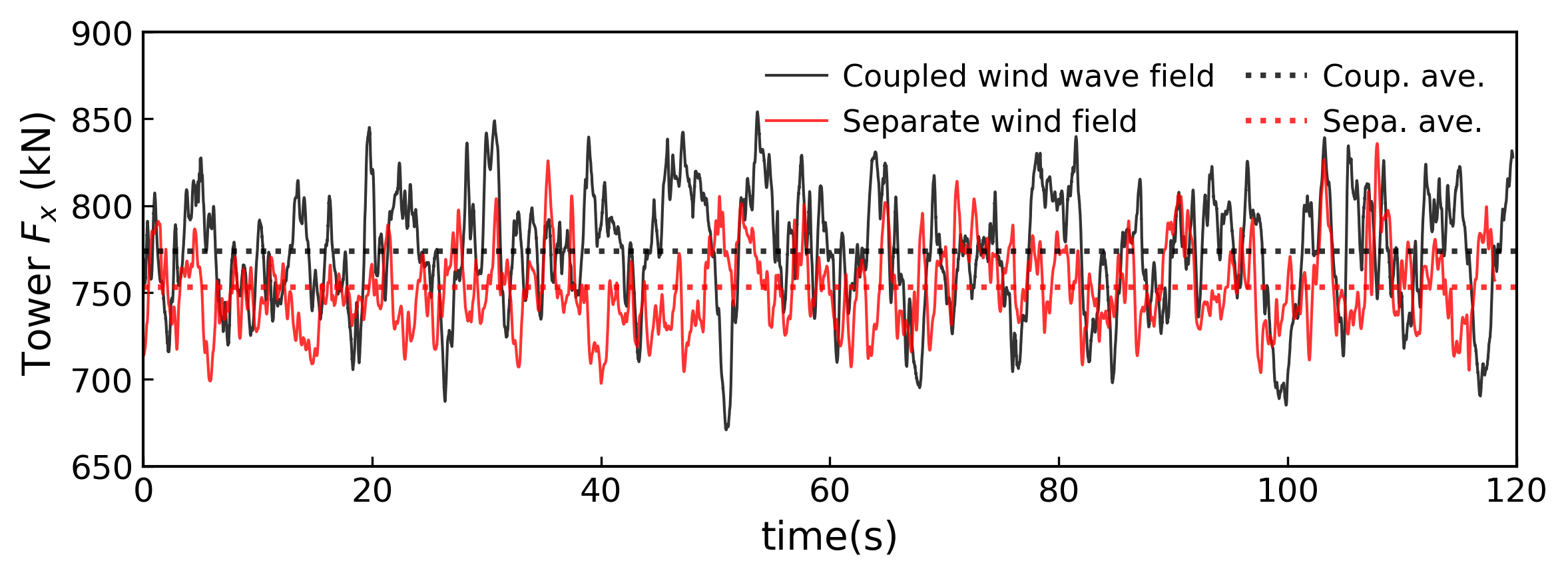}}
\subfigure[]{\label{fig:Axialforce_frequency_74}\includegraphics[width=0.8\textwidth]{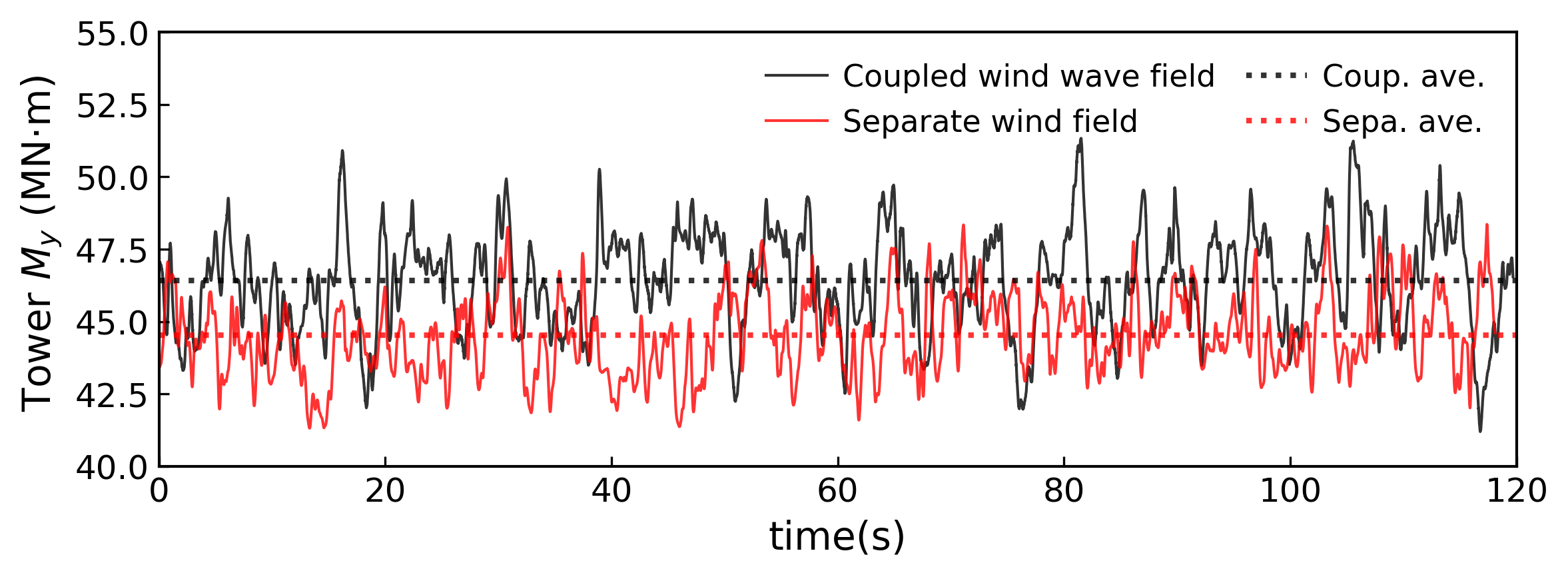}}
\caption{(a) Time history of the shear force $F_x$ at the bottom of the tower; (b) Time history of the bending moment $M_y$ at the bottom of the tower.}
\label{fig:ExtremePitch_Tower}

\end{figure}

The shear force and bending moment at the tower bottom also exhibit higher average values and greater variances under the coupled wind and wave fields, as depicted in Fig. \ref{fig:ExtremePitch_Tower}. A summary of the loadings applied at the nacelle, tower bottom, and monopile bottom is presented in Table \ref{table:ExtremePitchSummary}. The wind wave coupled effect results in an increase in the average $F_x^{to}$ and average $M_y^{to}$ by $2.9\%$ and $4.04\%$ respectively. It also leads to an increase in the maximum $F_x^{to}$ and maximum $M_y^{to}$ by $2.2\%$ and $6.15\%$ respectively. Additionally, the standard deviation of $F_x^{to}$ and $M_y^{to}$ is increased by $45.3\%$ and $26.7\%$ respectively. These findings highlight the impact of coupled wind and wave fields on the loadings experienced by the wind turbine structure.

Comparing the loading with a pitch angle of $90^\circ$ in Table \ref{table:ExtremeSummary} to the loading with a pitch angle of $60^\circ$ in Table \ref{table:ExtremePitchSummary}, significant differences can be observed. The maximum shear force at the bottom of the tower ($F_x^{to}$) increases from 613.1 kN to 854.1 kN, and the maximum bending moment ($M_y^{to}$) increases from 31.02 MNm to 51.3 MNm with the decrease of the pitch angle. The failed control of the pitch angle at $60^\circ$ results in a $65\%$ increase in the bending moment, which will exceeds the material limits and potentially leading to the failure of the tower with a worse control of pitch angle. These findings highlight the increased loading on the tower under extreme wind conditions with a pitch angle of $60^\circ$, indicating the importance of maintaining appropriate pitch control to ensure the structural integrity and safety of the wind turbine.

\begin{table}[h]
\centering
\caption{Normal forces at blade roots under extreme wind and wave conditions with a pitch angle of $60^{\circ}$}
\label{table:ExtremePitch_NormalForces}
\begin{tabular}{l | l l l|l l l|l l l}
\hline
\multicolumn{1}{c|}{$F_N$} & \multicolumn{3}{c|}{Average (kN)} & \multicolumn{3}{c|}{Maximum}&\multicolumn{3}{c}{RMS (kN)}\\
\hline
{No.} &coup.& sepa.&diff.&coup.& sepa.&diff. &coup.& sepa.&diff.\\
\hline
1 & 129.4  & 124.7 & 3.75$\%$ &153.8& 147.0 & 4.63$\%$&7.36& 6.60 & 11.61$\%$\\
2 & 90.8  & 86.4 & 5.15$\%$ &108.3& 100.6 & 7.67$\%$&5.03&4.76 & 5.59$\%$\\
3 & 162.3  & 154.6 &  4.9$\%$ &188.2& 182.8 & 2.93$\%$&8.99& 7.97& 12.88$\%$\\
\hline
\multicolumn{1}{c|}{$F_T$} & \multicolumn{3}{c|}{Average (kN)} & \multicolumn{3}{c|}{Maximum}&\multicolumn{3}{c}{RMS (kN)}\\
\hline
1 & 443.9  & 426.9 & 3.98$\%$ &530.4& 567.3 & -6.49$\%$&28.17& 26.80 & 5.12$\%$\\
2 & 432.1  & 410.6 & 5.22$\%$ &532.1& 492.5 & 8.04$\%$&30.72&30.77 & -0.16$\%$\\
3 & 388.8  & 373.2 &  4.16$\%$ &459.1& 466.6 & -1.61$\%$&25.82& 22.90& 12.74$\%$\\
\hline
\end{tabular}
\end{table}

\begin{table}[h]
\centering
\caption{Bending moment at blade roots under extreme conditions with a pitch angle of $60^{\circ}$}
\label{table:ExtremePitch_BladeBendingMoment}
\begin{tabular}{l | l l l| l l l| l l l}
\hline
\multicolumn{1}{c|}{$M_{out}$} & \multicolumn{3}{c|}{Average (MNm)} & \multicolumn{3}{c|}{Maximum (MNm)} & \multicolumn{3}{c}{RMS (MNm)}\\
\hline
{No.} &Coupled& sepa.&diff. &Coupled& sepa.&diff.&Coupled& sepa.&diff.\\
\hline
1 & 3.27  & 3.15 & 3.65$\%$ & 3.77 &3.78& -0.38$\%$ & 0.186 &0.167& 11.17$\%$\\
2 & 2.17 & 2.05& 5.57$\%$ & 2.68 &2.43& 10.28$\%$ & 0.131 &0.120& 8.45$\%$\\
3 & 4.07 & 3.88& 4.83$\%$ & 4.76 &4.71& 1.18$\%$ &0.235 & 0.209&12.21$\%$\\
\hline
\multicolumn{1}{c|}{$M_{in}$} & \multicolumn{3}{c|}{Average (MNm)} & \multicolumn{3}{c|}{Maximum (MNm)} & \multicolumn{3}{c}{RMS (MNm)}\\
\hline
1 & 12.06 & 11.55& 4.38$\%$ &14.56&15.28& 4.66$\%$ & 0.779 & 0.725&7.39$\%$\\
2 & 12.07  &11.41  & 5.82$\%$ & 14.44 &13.36& 8.07$\%$ & 0.733 & 0.735&-0.24$\%$\\
3 & 10.06  & 9.70& 3.72$\%$ & 12.20&12.18& 0.19$\%$ & 0.677& 0.630&7.50$\%$\\
\hline
\end{tabular}
\end{table}

\begin{table}[h]
\centering
\caption{Summary of the loading under extreme conditions with blade pitch of 60$^{\circ}$}
\label{table:ExtremePitchSummary}
\begin{tabular}{l | l l l| l l l| l l l}
\hline
\multicolumn{1}{c|}{} & \multicolumn{3}{c|}{Average} & \multicolumn{3}{c|}{Maximum} & \multicolumn{3}{c}{RMS}\\
\hline
{} &Coup.& sepa.&diff. &Coup.& sepa.&diff.&Coup.& sepa.&diff.\\
\hline
rotor $M_t^{ro}$ (MNm) & 34.35  & 32.82 & 4.64$\%$ & 38.36 &36.90& 3.97$\%$ &1.35 & 1.18&14.89$\%$\\
nacelle $F_x^{na}$ (kN) & 382.4  & 365.6 & 4.60$\%$ & 426.7&400.5& 6.53$\%$ & 14.25 &11.74& 21.42$\%$\\
nacelle $F_y^{na}$ (kN) & -32.68  & -32.21& 1.46$\%$ & -130.29&-113.75& 14.53$\%$ & 31.39 &30.88& 1.63$\%$\\

\hline
tower $F_x^{to}$ (kN) & 774.2  & 752.4 & 2.90$\%$ &854.1 &835.8& 2.19$\%$ & 33.41 & 23.00&45.28$\%$\\
tower $F_y^{to}$ (kN) & -35.52  & -34.92 & 1.74$\%$ & -145.7 &-121.6& 19.7$\%$ & 34.15 & 32.21&6.03$\%$\\
tower $M_x^{to}$ (MNm) & -2.39&-2.34 &2.24$\%$ & -9.75 &-9.00& 8.32$\%$ & 2.44 & 2.43&0.64$\%$\\
tower $M_y^{to}$ (MNm) & 46.40  & 44.60 & 4.04$\%$ & 51.33 &48.36& 6.15$\%$ & 1.72 & 1.36&26.7$\%$\\
\hline
monopile $F_x^{mo}$ (kN) & 962.6  & 896.2 & 7.41$\%$ &3701.9 &3664.0& 1.03$\%$ & 1802.5 & 1795.0&0.40$\%$\\
monopile $M_y^{mo}$ (MNm) & 86.68  & 80.84& 7.23$\%$ & 151.2 &141.5&6.82$\%$ & 33.42 & 33.49&-0.20$\%$\\
\hline
\end{tabular}
\end{table}

\subsubsection{Cut-off state with pitch angle of 0$^{\circ}$}

In the most severe scenario, where no pitch control is applied and the pitch angle remains fixed at $0^{\circ}$ (referred to as case 4), the simulations of wind and wave loadings on wind turbines are simulated. Tables \ref{table:ExtremePitchZero_NormalForces} and \ref{table:ExtremePitchZero_BladeBendingMoment} present the shear forces and bending moments at the blade roots. These tables reveal that the coupled effect of wind and waves significantly amplifies both the average and fluctuating loadings on the structural components. Specifically, the wind-wave interaction increase the maximum normal force at the blade root by 6.47$\%$, and the maximum out-of-plane bending moment by 5.76$\%$. As the pitch angle decreases from $60^{\circ}$ to $0^{\circ}$, the maximum normal force experiences a substantial increase from 188.2 kN to 535.6 kN, and the maximum out-of-plane bending moment increases from 4.76 MNm to 14.51 MNm. Notably, when pitch control is entirely absent, the loadings applied to the blade roots surpass those experienced under operational conditions.

Table \ref{table:ExtremePitchZeroSummary} provides a summary of the loadings at the nacelle, tower base, and monopile base when the pitch angle is $0^{\circ}$. At this configuration, the maximum shear force and maximum bending moment at the tower base exceed 1863.2 kN and 132.1 MNm, respectively, with the maximum bending stress surpassing 527 MPa, exceeding the material's yield stress of 380 MPa. This suggests that the absence of pitch control during extreme wind conditions can potentially lead to the failure of the wind turbine tower. Moreover, the wind-wave interaction increases the maximum bending moment at the tower base by 8.69$\%$. Furthermore, at the monopile foundation, the maximum shear force and bending moment exceed 4679.9 kN and 261.4 MNm, respectively. These extreme loadings can also induce the failure of the monopile foundation.

\begin{table}[h]
\centering
\caption{Normal forces at blade roots under extreme wind and wave conditions with a pitch angle of $0^{\circ}$}
\label{table:ExtremePitchZero_NormalForces}
\begin{tabular}{l | l l l|l l l|l l l}
\hline
\multicolumn{1}{c|}{$F_N$} & \multicolumn{3}{c|}{Average (kN)} & \multicolumn{3}{c|}{Maximum}&\multicolumn{3}{c}{RMS (kN)}\\
\hline
{Blade No.} &coup.& sepa.&diff.&coup.& sepa.&diff. &coup.& sepa.&diff.\\
\hline
1 & 450.4  & 432.8 & 4.04$\%$ &535.6& 503.1 & 6.47$\%$&26.08& 20.65 & 26.27$\%$\\
2 & 428.3  & 406.5 & 5.36$\%$ &496.7&475.8 & 4.37$\%$&23.83&21.59 & 10.38$\%$\\
3 & 418.7  & 400.9 &  4.44$\%$ &496.9& 475.9 & 4.41$\%$&24.55& 23.50& 4.46$\%$\\
\hline
\multicolumn{1}{c|}{$F_T$} & \multicolumn{3}{c|}{Average (kN)} & \multicolumn{3}{c|}{Maximum}&\multicolumn{3}{c}{RMS (kN)}\\
\hline
1 & 87.8  & 84.3 & 4.17$\%$ &108.2& 100.5 & 7.64$\%$&6.17& 4.95 & 24.5$\%$\\
2 & 122.6  & 116.4 & 5.34$\%$ &144.1& 136.8 & 5.35$\%$&7.31&6.78 & 7.85$\%$\\
3 & 64.8  & 62.1 &  4.31$\%$ &78.3& 77.7 & 0.81$\%$&4.32& 4.19& 3.07$\%$\\
\hline
\end{tabular}
\end{table}

\begin{table}[h]
\centering
\caption{Bending moment at blade roots under extreme conditions with a pitch angle of $0^{\circ}$}
\label{table:ExtremePitchZero_BladeBendingMoment}
\begin{tabular}{l | l l l| l l l| l l l}
\hline
\multicolumn{1}{c|}{$M_{out}$} & \multicolumn{3}{c|}{Average (MNm)} & \multicolumn{3}{c|}{Maximum (MNm)} & \multicolumn{3}{c}{RMS (MNm)}\\
\hline
{Blade No.} &Coupled& separate&diff. &Coupled& separate&diff.&Coupled& separate&diff.\\
\hline
1 & 12.41  & 11.95 & 3.91$\%$ & 14.51 &14.28& 1.61$\%$ & 0.653 &0.552& 18.35$\%$\\
2 & 11.59 & 10.96& 5.76$\%$ & 13.34 &13.10& 1.88$\%$ & 0.638 &0.605& 5.60$\%$\\
3 & 11.53 & 11.02& 4.64$\%$ & 13.60 &12.83& 6.02$\%$ &0.677 & 0.631&7.26$\%$\\
\hline
\multicolumn{1}{c|}{$M_{in}$} & \multicolumn{3}{c|}{Average (MNm)} & \multicolumn{3}{c|}{Maximum (MNm)} & \multicolumn{3}{c}{RMS (MNm)}\\
\hline
1 & 2.05 & 1.97& 4.15$\%$ &2.48&2.29& 8.32$\%$ & 0.126 & 0.100&25.94$\%$\\
2 & 2.98  &2.82  & 5.61$\%$ & 3.47 &3.36& 3.19$\%$ & 0.167 & 0.158&5.43$\%$\\
3 & 1.53  & 1.47& 4.52$\%$ & 1.83&1.76& 3.76$\%$ & 0.093& 0.089&4.37$\%$\\
\hline
\end{tabular}
\end{table}

\begin{table}[h]
\centering
\caption{Summary of the loading under extreme conditions with blade picth of 0$^{\circ}$}
\label{table:ExtremePitchZeroSummary}
\begin{tabular}{l | l l l| l l l| l l l}
\hline
\multicolumn{1}{c|}{} & \multicolumn{3}{c|}{Average} & \multicolumn{3}{c|}{Maximum} & \multicolumn{3}{c}{RMS}\\
\hline
{} &Coup.& sepa.&diff. &Coup.& sepa.&diff.&Coup.& sepa.&diff.\\
\hline
rotor $M_t^{ro}$ (MNm) & 5.668  & 5.413 & 4.71$\%$ & 4.925 &4.849& 1.57$\%$ &0.239 &0.170&40.47$\%$\\
nacelle $F_x^{na}$ (kN) & 1297.5  & 1240.3 & 4.60$\%$ & 1470.3&1348.7& 9.02$\%$ & 51.58 &36.55& 41.14$\%$\\
nacelle $F_y^{na}$ (kN) & -4.90  & -4.74& 4.40$\%$ & -31.36&-20.54& 52.69$\%$ & 6.28 &5.98& 5.00$\%$\\

\hline
tower $F_x^{to}$ (kN) & 1692.9  & 1626.8 & 4.06$\%$ &1863.2 &1757.1& 6.04$\%$ & 64.01 & 42.24&51.53$\%$\\
tower $F_y^{to}$ (kN) & -2.79  & -1.53 & 82.85$\%$ &-42.03 &-41.87& 0.37$\%$ & 12.16 & 11.31&7.57$\%$\\
tower $M_x^{to}$ (MNm) & -0.548&-0.607 &-9.72$\%$ & -2.903 &-2.704& 7.37$\%$ & 0.704 & 0.656&7.25$\%$\\
tower $M_y^{to}$ (MNm) & 117.6  & 112.5 & 4.48$\%$ & 132.1 &121.6& 8.69$\%$ & 4.1 & 3.03&45.62$\%$\\
\hline
monopile $F_x^{mo}$ (kN) & 1867.4  & 1765.9 & 5.75$\%$ &4679.9 &4555.4& 2.73$\%$ & 1829.2 & 1795.7&1.86$\%$\\
monopile $M_y^{mo}$ (MNm) & 193.9  & 184.7& 5.01$\%$ & 261.4 &246.2&6.18$\%$ & 35.86 & 33.56&6.87$\%$\\
\hline
\end{tabular}
\end{table}

This section analyzes the wind and wave loading on an offshore wind turbine with a monopile foundation. We consider three cases: the wind turbine in operational condition, wind turbine during shut-down condition with successful pitch control, and wind turbine under shut-off condition with failed pitch control. The operational condition represents normal operation, where the wind turbine is actively generating power. The shut-down condition with successful pitch control corresponds to a situation where the turbine is intentionally shut down and the blades are adjusted to minimize wind loadings. Lastly, the shut-off condition with failed pitch control represents a scenario where the blade pitch control mechanism malfunctions, leading to inefficient control of wind loadings. For each case, the loadings induced by coupled wind and wave fields with those induced by separate wind and wave fields are compared. Under operational conditions, the loadings induced by coupled wind and wave fields are found to be similar to those induced by separate wind and wave fields. However, under extreme conditions characterized by high wave heights and strong wind forcing, the interaction between the wave surface and the wind introduces additional complexities. The wave induces wind turbulence, alters the average wind velocity, and affects the wind wave loading on the offshore wind turbine. This results in increased wind wave loading and more significant variability in the loading.

\section{Conclusion}

The wind and wave coupling effect affects the characteristics of wind fields and wave profiles. Due to the complex physics, it is challenging to accurately model the evolution of coupled wind and wave. Currently, researchers treat wind and wave loading on offshore structures separately. In addition, existing research on two-phase wind-wave modeling mainly analyzed the coupling effect under uniform wind forcing without considering the inherent strong turbulences of wind above the air-water interface. This is adequate to realistically and accurately quantify the combined wind-wave loading effects on coastal and offshore structures. 

The present study develops a two-phase flow model to simulate coupled turbulent wind-wave fields using the open source program OpenFOAM. The volume of fluid (VOF) method is adopted to capture the complex water-air interface configuration. An advection equation for the VOF indicator function is introduced to track the motion of two-phase flows. The wind-wave field conditions at the inlet boundary are defined through specifying water particle velocities, turbulent wind velocities and wave surface elevation. The two-phase model is utilized to analyze the coupled interaction of wind and wave on a 100 m scale especially under extreme wind and wave conditions. In addition, the coupled wind and wave fields are applied to analyze the combined wind and wave loadings acting on an offshore wind turbine. Based on the presented results, we can obtain the following conclusions.

\begin{enumerate}

  \item The location of the region of intense wind turbulence depends on the relative speed between wind and wave. The intense wind turbulence occurs at the upwind side of the wave crest surface if the wave travels faster than the wind or occurs at the downwind side if the wind travels faster. The wave-induced wind turbulence increases when the wind forcing velocity and/or wave height increases. Quantitatively, extreme wind forcing ($U_{10}$=55 m/s) can increase the wind turbulence by over 100$\%$.
  
  \item Different relative speeds between wind and wave can induce different positive-negative patterns of wave coherent velocities. When the wave travels faster, the wave coherent velocities are mainly induced by the movement of waves and are confined very close to the air-water interface. When wind travels faster, the wave coherent velocities are caused by the fast-moving wind.

  \item Under wind turbine operational conditions, characterized by a wind speed of $U_{10}=10$ m/s and a wave height of 4 m, the presence of waves affects the wind turbulence very close to the wave surface. The wind-wave coupling effect on the combined wind-wave loadings acting on the offshore wind turbine is slight. The difference in average and maximum loadings at key locations, such as the nacelle, the tower bottom, and the monopile bottom, is within $3\%$. 
  
  \item Under extreme conditions, characterized by a high wind speed of $U_{10}=55$ m/s and a significant wave height of 10 m, the presence of waves has a noticeable impact on the wind structure within a region extending up to approximately 45 m in height. In this scenario, the coupled wind and wave fields lead to an increase in the average aerodynamic loading and a significant amplification of the fluctuation in the aerodynamic loading compared to the loading under separate wind and wave conditions. Specifically, the maximum out-of-plane shear force at the blade roots experiences an increase of approximately $10\%$. The maximum bending moment $F_y$ at both the tower bottom and the monopile bottom also experiences an increase of around $6\%$, suggesting a greater bending moment on these components. Furthermore, the effect of the coupled wind and wave fields is evident in the standard deviation of the aerodynamic loading at the tower bottom. The standard deviation of the shear force at the tower bottom increases by up to approximately $45\%$. Similarly, the standard deviation of the bending moment at the tower bottom increases by approximately $27\%$, highlighting the increased variability in the loading. Under coupled wind wave fields, the spectrum of the loadings has a significant peak at the frequency of wave frequency.

  \item In the presence of extreme wind and wave conditions, characterized by a high wind speed of $U_{10}=55$ m/s and a significant wave height of 10 m, the absence of pitch control can potentially lead to the failure of the tower. Specifically, when the pitch angle is set to $0^{\circ}$, the bending stress at the tower's base exceeds its yield stress threshold.
  
\end{enumerate}

In summary, the developed two-phase model for simulating turbulent wind-wave fields is utilized to analyze the coupling effect between wind and wave under extreme conditions and applied to analyze the coupled wind and wave loading on offshore wind turbines. The developed model and presented results advance the understanding on realistically coupled wind-wave field characteristics and will serve as the foundation for research on critical infrastructure systems exposed to wind-wave impacts. Under typical operational conditions, the influence of wave-induced turbulence on the wind and wave loadings is relatively minor. The small differences in loading indicate that the combined effect of wind and waves does not significantly impact the turbine's performance or pose any immediate concerns in terms of safety or operational limitations. It is important to note that the specific loading characteristics vary depending on the wind and wave conditions. Under extreme wind conditions, the results emphasize the significant influence of the combined wind and wave conditions on the aerodynamic loading on the offshore wind turbine. The amplified fluctuations and increased average loading values indicate the need for careful consideration of the combined effects of wind and waves in the design, operation, and maintenance of offshore wind turbines to ensure their structural integrity and performance in extreme environments.

\newpage
\section*{Acknowledgement}

This work was supported by the Louisiana State University Economic Development Assistantship and the research was conducted using high performance computing resources provided by Louisiana State University. The authors are grateful for all the support. 
\newpage
\clearpage

\newpage

\newpage
\bibliographystyle{elsarticle-num-names}
\bibliography{main.bib}

\begin{thebibliography}{28}
\expandafter\ifx\csname natexlab\endcsname\relax\def\natexlab#1{#1}\fi
\providecommand{\url}[1]{\texttt{#1}}
\providecommand{\href}[2]{#2}
\providecommand{\path}[1]{#1}
\providecommand{\DOIprefix}{doi:}
\providecommand{\ArXivprefix}{arXiv:}
\providecommand{\URLprefix}{URL: }
\providecommand{\Pubmedprefix}{pmid:}
\providecommand{\doi}[1]{\href{http://dx.doi.org/#1}{\path{#1}}}
\providecommand{\Pubmed}[1]{\href{pmid:#1}{\path{#1}}}
\providecommand{\bibinfo}[2]{#2}
\ifx\xfnm\relax \def\xfnm[#1]{\unskip,\space#1}\fi
\bibitem[{Sun and Jahangiri(2018)}]{sun2018bi}
\bibinfo{author}{C.~Sun}, \bibinfo{author}{V.~Jahangiri},
\newblock \bibinfo{title}{Bi-directional vibration control of offshore wind turbines using a 3d pendulum tuned mass damper},
\newblock \bibinfo{journal}{Mechanical Systems and Signal Processing} \bibinfo{volume}{105} (\bibinfo{year}{2018}) \bibinfo{pages}{338--360}.
\bibitem[{Li et~al.(2018)Li, Liu, Yuan, and Gao}]{li2018wind}
\bibinfo{author}{L.~Li}, \bibinfo{author}{Y.~Liu}, \bibinfo{author}{Z.~Yuan}, \bibinfo{author}{Y.~Gao},
\newblock \bibinfo{title}{Wind field effect on the power generation and aerodynamic performance of offshore floating wind turbines},
\newblock \bibinfo{journal}{Energy} \bibinfo{volume}{157} (\bibinfo{year}{2018}) \bibinfo{pages}{379--390}.
\bibitem[{Doubrawa et~al.(2019)Doubrawa, Churchfield, Godvik, and Sirnivas}]{doubrawa2019load}
\bibinfo{author}{P.~Doubrawa}, \bibinfo{author}{M.~J. Churchfield}, \bibinfo{author}{M.~Godvik}, \bibinfo{author}{S.~Sirnivas},
\newblock \bibinfo{title}{Load response of a floating wind turbine to turbulent atmospheric flow},
\newblock \bibinfo{journal}{Applied Energy} \bibinfo{volume}{242} (\bibinfo{year}{2019}) \bibinfo{pages}{1588--1599}.
\bibitem[{Feyzollahzadeh et~al.(2016)Feyzollahzadeh, Mahmoodi, Yadavar-Nikravesh, and Jamali}]{feyzollahzadeh2016wind}
\bibinfo{author}{M.~Feyzollahzadeh}, \bibinfo{author}{M.~Mahmoodi}, \bibinfo{author}{S.~Yadavar-Nikravesh}, \bibinfo{author}{J.~Jamali},
\newblock \bibinfo{title}{Wind load response of offshore wind turbine towers with fixed monopile platform},
\newblock \bibinfo{journal}{Journal of Wind Engineering and Industrial Aerodynamics} \bibinfo{volume}{158} (\bibinfo{year}{2016}) \bibinfo{pages}{122--138}.
\bibitem[{Rendon and Manuel(2014)}]{rendon2014long}
\bibinfo{author}{E.~A. Rendon}, \bibinfo{author}{L.~Manuel},
\newblock \bibinfo{title}{Long-term loads for a monopile-supported offshore wind turbine},
\newblock \bibinfo{journal}{Wind Energy} \bibinfo{volume}{17} (\bibinfo{year}{2014}) \bibinfo{pages}{209--223}.
\bibitem[{Oh et~al.(2013)Oh, Kim, and Lee}]{oh2013preliminary}
\bibinfo{author}{K.-Y. Oh}, \bibinfo{author}{J.-Y. Kim}, \bibinfo{author}{J.-S. Lee},
\newblock \bibinfo{title}{Preliminary evaluation of monopile foundation dimensions for an offshore wind turbine by analyzing hydrodynamic load in the frequency domain},
\newblock \bibinfo{journal}{Renewable energy} \bibinfo{volume}{54} (\bibinfo{year}{2013}) \bibinfo{pages}{211--218}.
\bibitem[{Winter et~al.(2020)Winter, Alam, Shekhar, Motley, Eberhard, Barbosa, Lomonaco, Arduino, and Cox}]{winter2020tsunami}
\bibinfo{author}{A.~O. Winter}, \bibinfo{author}{M.~S. Alam}, \bibinfo{author}{K.~Shekhar}, \bibinfo{author}{M.~R. Motley}, \bibinfo{author}{M.~O. Eberhard}, \bibinfo{author}{A.~R. Barbosa}, \bibinfo{author}{P.~Lomonaco}, \bibinfo{author}{P.~Arduino}, \bibinfo{author}{D.~T. Cox},
\newblock \bibinfo{title}{Tsunami-like wave forces on an elevated coastal structure: Effects of flow shielding and channeling},
\newblock \bibinfo{journal}{Journal of Waterway, Port, Coastal, and Ocean Engineering} \bibinfo{volume}{146} (\bibinfo{year}{2020}) \bibinfo{pages}{04020021}.
\bibitem[{Wienke and Oumeraci(2005)}]{wienke2005breaking}
\bibinfo{author}{J.~Wienke}, \bibinfo{author}{H.~Oumeraci},
\newblock \bibinfo{title}{Breaking wave impact force on a vertical and inclined slender pile—theoretical and large-scale model investigations},
\newblock \bibinfo{journal}{Coastal engineering} \bibinfo{volume}{52} (\bibinfo{year}{2005}) \bibinfo{pages}{435--462}.
\bibitem[{Bobillier et~al.(2001)Bobillier, Chakrabarti, and Christiansen}]{bobillier2001physical}
\bibinfo{author}{B.~Bobillier}, \bibinfo{author}{S.~Chakrabarti}, \bibinfo{author}{P.~Christiansen},
\newblock \bibinfo{title}{Physical modeling of wind load on a floating offshore structure},
\newblock \bibinfo{journal}{J. Offshore Mech. Arct. Eng.} \bibinfo{volume}{123} (\bibinfo{year}{2001}) \bibinfo{pages}{170--176}.
\bibitem[{Buljac et~al.(2022)Buljac, Kozmar, Yang, and Kareem}]{buljac2022concurrent}
\bibinfo{author}{A.~Buljac}, \bibinfo{author}{H.~Kozmar}, \bibinfo{author}{W.~Yang}, \bibinfo{author}{A.~Kareem},
\newblock \bibinfo{title}{Concurrent wind, wave and current loads on a monopile-supported offshore wind turbine},
\newblock \bibinfo{journal}{Engineering Structures} \bibinfo{volume}{255} (\bibinfo{year}{2022}) \bibinfo{pages}{113950}.
\bibitem[{Li et~al.(2015)Li, Castro, Sinokrot, Prescott, and Carrica}]{li2015coupled}
\bibinfo{author}{Y.~Li}, \bibinfo{author}{A.~Castro}, \bibinfo{author}{T.~Sinokrot}, \bibinfo{author}{W.~Prescott}, \bibinfo{author}{P.~Carrica},
\newblock \bibinfo{title}{Coupled multi-body dynamics and cfd for wind turbine simulation including explicit wind turbulence},
\newblock \bibinfo{journal}{Renewable Energy} \bibinfo{volume}{76} (\bibinfo{year}{2015}) \bibinfo{pages}{338--361}.
\bibitem[{Zhou et~al.(2022)Zhou, Xiao, Liu, Incecik, Peyrard, Wan, Pan, and Li}]{zhou2022exploring}
\bibinfo{author}{Y.~Zhou}, \bibinfo{author}{Q.~Xiao}, \bibinfo{author}{Y.~Liu}, \bibinfo{author}{A.~Incecik}, \bibinfo{author}{C.~Peyrard}, \bibinfo{author}{D.~Wan}, \bibinfo{author}{G.~Pan}, \bibinfo{author}{S.~Li},
\newblock \bibinfo{title}{Exploring inflow wind condition on floating offshore wind turbine aerodynamic characterisation and platform motion prediction using blade resolved cfd simulation},
\newblock \bibinfo{journal}{Renewable Energy} \bibinfo{volume}{182} (\bibinfo{year}{2022}) \bibinfo{pages}{1060--1079}.
\bibitem[{Nicoud and Ducros(1999)}]{nicoud1999subgrid}
\bibinfo{author}{F.~Nicoud}, \bibinfo{author}{F.~Ducros},
\newblock \bibinfo{title}{Subgrid-scale stress modelling based on the square of the velocity gradient tensor},
\newblock \bibinfo{journal}{Flow, turbulence and Combustion} \bibinfo{volume}{62} (\bibinfo{year}{1999}) \bibinfo{pages}{183--200}.
\bibitem[{Ben-Nasr et~al.(2017)Ben-Nasr, Hadjadj, Chaudhuri, and Shadloo}]{ben2017assessment}
\bibinfo{author}{O.~Ben-Nasr}, \bibinfo{author}{A.~Hadjadj}, \bibinfo{author}{A.~Chaudhuri}, \bibinfo{author}{M.~Shadloo},
\newblock \bibinfo{title}{Assessment of subgrid-scale modeling for large-eddy simulation of a spatially-evolving compressible turbulent boundary layer},
\newblock \bibinfo{journal}{Computers \& Fluids} \bibinfo{volume}{151} (\bibinfo{year}{2017}) \bibinfo{pages}{144--158}.
\bibitem[{Weickert et~al.(2010)Weickert, Teike, Schmidt, and Sommerfeld}]{weickert2010investigation}
\bibinfo{author}{M.~Weickert}, \bibinfo{author}{G.~Teike}, \bibinfo{author}{O.~Schmidt}, \bibinfo{author}{M.~Sommerfeld},
\newblock \bibinfo{title}{Investigation of the les wale turbulence model within the lattice boltzmann framework},
\newblock \bibinfo{journal}{Computers \& Mathematics with Applications} \bibinfo{volume}{59} (\bibinfo{year}{2010}) \bibinfo{pages}{2200--2214}.
\bibitem[{Higuera(2017)}]{olaFlow}
\bibinfo{author}{P.~Higuera}, \bibinfo{title}{olaflow: {CFD} for waves [{S}oftware].}, \bibinfo{year}{2017}. \URLprefix \url{https://doi.org/10.5281/zenodo.1297013}. \DOIprefix\doi{10.5281/zenodo.1297013}.
\bibitem[{Kr{\"o}ger and Kornev(2018)}]{kroger2018generation}
\bibinfo{author}{H.~Kr{\"o}ger}, \bibinfo{author}{N.~Kornev},
\newblock \bibinfo{title}{Generation of divergence free synthetic inflow turbulence with arbitrary anisotropy},
\newblock \bibinfo{journal}{Computers \& Fluids} \bibinfo{volume}{165} (\bibinfo{year}{2018}) \bibinfo{pages}{78--88}.
\bibitem[{Poletto et~al.(2013)Poletto, Craft, and Revell}]{poletto2013new}
\bibinfo{author}{R.~Poletto}, \bibinfo{author}{T.~Craft}, \bibinfo{author}{A.~Revell},
\newblock \bibinfo{title}{A new divergence free synthetic eddy method for the reproduction of inlet flow conditions for les},
\newblock \bibinfo{journal}{Flow, turbulence and combustion} \bibinfo{volume}{91} (\bibinfo{year}{2013}) \bibinfo{pages}{519--539}.
\bibitem[{Kornev and Hassel(2007)}]{kornev2007synthesis}
\bibinfo{author}{N.~Kornev}, \bibinfo{author}{E.~Hassel},
\newblock \bibinfo{title}{Synthesis of homogeneous anisotropic divergence-free turbulent fields with prescribed second-order statistics by vortex dipoles},
\newblock \bibinfo{journal}{Physics of Fluids} \bibinfo{volume}{19} (\bibinfo{year}{2007}) \bibinfo{pages}{068101}.
\bibitem[{Jacobsen et~al.(2012)Jacobsen, Fuhrman, and Freds{\o}e}]{jacobsen2012wave}
\bibinfo{author}{N.~G. Jacobsen}, \bibinfo{author}{D.~R. Fuhrman}, \bibinfo{author}{J.~Freds{\o}e},
\newblock \bibinfo{title}{A wave generation toolbox for the open-source cfd library: Openfoam{\textregistered}},
\newblock \bibinfo{journal}{International Journal for numerical methods in fluids} \bibinfo{volume}{70} (\bibinfo{year}{2012}) \bibinfo{pages}{1073--1088}.
\bibitem[{Fuhrman et~al.(2006)Fuhrman, Madsen, and Bingham}]{fuhrman2006numerical}
\bibinfo{author}{D.~R. Fuhrman}, \bibinfo{author}{P.~A. Madsen}, \bibinfo{author}{H.~B. Bingham},
\newblock \bibinfo{title}{Numerical simulation of lowest-order short-crested wave instabilities},
\newblock \bibinfo{journal}{Journal of Fluid Mechanics} \bibinfo{volume}{563} (\bibinfo{year}{2006}) \bibinfo{pages}{415}.
\bibitem[{Troldborg(2009)}]{troldborg2009actuator}
\bibinfo{author}{N.~Troldborg},
\newblock \bibinfo{title}{Actuator line modeling of wind turbine wakes}  (\bibinfo{year}{2009}).
\bibitem[{Weller et~al.(1998)Weller, Tabor, Jasak, and Fureby}]{weller1998tensorial}
\bibinfo{author}{H.~G. Weller}, \bibinfo{author}{G.~Tabor}, \bibinfo{author}{H.~Jasak}, \bibinfo{author}{C.~Fureby},
\newblock \bibinfo{title}{A tensorial approach to computational continuum mechanics using object-oriented techniques},
\newblock \bibinfo{journal}{Computers in physics} \bibinfo{volume}{12} (\bibinfo{year}{1998}) \bibinfo{pages}{620--631}.
\bibitem[{Habchi et~al.(2013)Habchi, Russeil, Bougeard, Harion, Lemenand, Ghanem, Della~Valle, and Peerhossaini}]{habchi2013partitioned}
\bibinfo{author}{C.~Habchi}, \bibinfo{author}{S.~Russeil}, \bibinfo{author}{D.~Bougeard}, \bibinfo{author}{J.-L. Harion}, \bibinfo{author}{T.~Lemenand}, \bibinfo{author}{A.~Ghanem}, \bibinfo{author}{D.~Della~Valle}, \bibinfo{author}{H.~Peerhossaini},
\newblock \bibinfo{title}{Partitioned solver for strongly coupled fluid--structure interaction},
\newblock \bibinfo{journal}{Computers \& Fluids} \bibinfo{volume}{71} (\bibinfo{year}{2013}) \bibinfo{pages}{306--319}.
\bibitem[{Holzmann(2016)}]{holzmann2016mathematics}
\bibinfo{author}{T.~Holzmann},
\newblock \bibinfo{title}{Mathematics, numerics, derivations and openfoam{\textregistered}},
\newblock \bibinfo{journal}{Loeben, Germany: Holzmann CFD}  (\bibinfo{year}{2016}).
\bibitem[{Benjamin(1959)}]{benjamin1959shearing}
\bibinfo{author}{T.~B. Benjamin},
\newblock \bibinfo{title}{Shearing flow over a wavy boundary},
\newblock \bibinfo{journal}{Journal of Fluid Mechanics} \bibinfo{volume}{6} (\bibinfo{year}{1959}) \bibinfo{pages}{161--205}.
\bibitem[{85020(2001)}]{esdu2001characteristics}
\bibinfo{author}{E.~85020}, \bibinfo{title}{Characteristics of atmospheric turbulence near the ground, part ii: single point data for strong winds (neutral atmosphere)}, \bibinfo{year}{2001}.
\bibitem[{Buckley and Veron(2016)}]{buckley2016structure}
\bibinfo{author}{M.~P. Buckley}, \bibinfo{author}{F.~Veron},
\newblock \bibinfo{title}{Structure of the airflow above surface waves},
\newblock \bibinfo{journal}{Journal of Physical Oceanography} \bibinfo{volume}{46} (\bibinfo{year}{2016}) \bibinfo{pages}{1377--1397}.

\end{thebibliography}

\end{document}